\documentclass[12pt,preprint]{aastex}
\usepackage{amssymb, amsmath, fullpage, graphicx,bm}
\shorttitle{RCB stars from double degenerate white dwarf mergers}
\shortauthors{Staff et al.}

\begin{document}
\title{Do R Coronae Borealis Stars Form from Double White Dwarf Mergers?}
\author{Jan. E. Staff}
\affil{Department of Physics and Astronomy, Louisiana State University,
202 Nicholson Hall, Tower Dr., Baton Rouge, LA 70803-4001, 
USA}
\author{Athira Menon\footnote{NuGrid collaboration} and Falk Herwig$^1$}
\affil{Department of Physics \& Astronomy, University of Victoria, Victoria,
BC V8P5C2, Canada}
\author{Wesley Even and Chris L. Fryer$^1$}
\affil{Los Alamos National Laboratory, Los Alamos, NM 87545, USA}
\author{Patrick M. Motl}
\affil{Department of Science, Mathematics \& Informatics,
Indiana University Kokomo, Kokomo, Indiana 46904-9003, USA}
\author{Tom Geballe}
\affil{ Gemini Observatory, 670 North A'ohoku Place, Hilo, HI 96720, USA}
\author{Marco Pignatari$^1$}
\affil{Department of Physics, University of Basel,
Klingelbergstrasse 82, CH-4056 Basel, Switzerland}
\author{Geoffrey C. Clayton and Joel E. Tohline}
\affil{Department of Physics and Astronomy, Louisiana State University,
202 Nicholson Hall, Tower Dr., Baton Rouge, LA 70803-4001,
USA}

\date{}

%
%

\begin{abstract} 

A leading formation scenario for R Coronae Borealis (RCB) stars invokes
the merger of degenerate He and CO white dwarfs (WD) in a binary.  The
observed ratio of $^{16}\mathrm{O}$/$^{18}\mathrm{O}$ for RCB stars is in
the range of 0.3-20 much smaller than the solar value of $\sim 500$.  In
this paper, we investigate whether such a low ratio can be obtained in
simulations of the merger of a CO and a He white dwarf.  We present the
results of five 3-dimensional hydrodynamic simulations of the merger of a
double white dwarf system where the total mass is $0.9 M_\odot$ and the
initial mass ratio (q) varies between 0.5 and 0.99.  We identify in
simulations with $q\lesssim0.7$ a feature around the merged stars where the
temperatures and densities are suitable for forming $^{18}\mathrm{O}$. 
However, more $^{16}\mathrm{O}$ is being dredged-up from the C- and O-rich
accretor during the merger than the amount of $^{18}\mathrm{O}$ that is
produced.  Therefore, on a dynamical time scale over which our hydrodynamics
simulation runs, a $^{16}\mathrm{O}/^{18}\mathrm{O}$ ratio of $\sim 2000$ in
the ``best'' case is found.  If the conditions found in the hydrodynamic
simulations persist for $10^6$ seconds the oxygen ratio drops to 16 in
one case studied, while in a hundred years it drops to $\sim 4$ in another
case studied, consistent with the observed values in RCB stars. 
Therefore, the merger of two white dwarfs remains a strong candidate for the
formation of these enigmatic stars.

\end{abstract}

%
%
\section{Introduction}

R Coronae Borealis stars (RCBs) are hydrogen deficient stars, with a carbon
rich atmosphere \citep{clayton96,clayton12}.  These very unusual stars are observed to
be approximately $98\%$ He and $1\%$ C by mass.  The masses of RCB stars are
difficult to measure since they have never been observed in a binary system,
but stellar pulsation models have shown masses to be on the order of $1
M_\odot$ \citep{saio08,han98}.  The luminosity is characterized by a
peculiar behavior: they fade at irregular intervals by up to 8 magnitudes,
and gradually recover back to maximum luminosity over a period of a few
months to a year. Such an observational feature is thought to be caused by
clouds of carbon dust formed by the star itself \citep{okeefe39}.

RCB stars show many anomalous elemental abundances compared to solar.
Typically they are extremely deficient in hydrogen and are enriched relative
to Fe, in N, Al, Na, Si, S, Ni, the s-process elements, and sometimes O
\citep{asplund00}.  The lower bound on the $^{12}$C/$^{13}$C ratio is 
between 14-100 for the majority of RCB stars, much larger than the 
equilibrium value in stars of solar metallicity which is 3.4 \citep{hema12}, 
although at least one star, V CrA, shows a significant abundance of $^{13}$C 
\citep{rao08,asplund00}.  Also, lithium has been detected in 5 RCB stars
\citep{asplund00,kipper06}.  
In other RCBs there is no lithium observed.
The atmospheres of these stars show material
processed during H burning via the CNO cycle and He burning via the 3-$\alpha$
process.  In this paper, we focus on the more recent discovery of the oxygen
isotopic ratio, $^{16}\mathrm{O}$ to $^{18}\mathrm{O}$
\citep{clayton07,garcia09,garcia10}, found to be of order unity in RCB stars
(the stars measured had ratios between 0.3 and 20).  
This ratio is found to be $\sim$ 500 in the
solar neighborhood \citep{scott06}, and varies from 200 to 600 in the
Galactic interstellar medium \citep{wilson94}.  No other known class of
stars displays $^{16}\mathrm{O}/^{18}\mathrm{O}\sim$1
\citep{clayton07}.

In a single star, partial He burning on the cool edge of the He burning shell  produces a significant amount of $^{18}$O but normally it would not be mixed to the surface. If He burning continues to its conclusion the $^{18}$O will be turned into $^{22}$Ne \citep{clayton05}.  Two scenarios have been put forth to explain
the progenitor evolution for RCB stars; one is a final helium shell flash
and the other a double degenerate white dwarf (WD) merger
\citep{webbink84,renzini90}.  According to \citet{iben96}, RCB stars could
be the result of a final flash, when a single star late in its evolution has
left the asymptotic giant branch and is cooling to form a WD
and a shell of helium surrounding the core ignites.  However, the
temperatures that result from He burning in a final flash will result in
$^{14}$N being completely burned into $^{22}$Ne leaving little
$^{18}\mathrm{O}$ \citep{clayton07}.

In the second scenario, a close binary system consisting of a He and a CO WD
merges, leading to an RCB star \citep{webbink84}.  
\citet{iben96} explain that theoretically the accretion of a He WD 
$\sim 0.3 M_{\odot}$ onto a CO WD $\sim0.6 M_{\odot}$ can
produce a carbon-rich supergiant star (M$\sim$0.9
M$_{\sun}$) that is hydrogen-deficient at the surface after two common envelope (CE) phases.  In this scenario, the He WD is disrupted and forms the envelope of the newly merged star while
the CO WD forms the core.  In He burning conditions the partial completion of the reaction chain ($^{14}\mathrm{N(\alpha,\gamma)}^{18}\mathrm{F(\beta^+)}
^{18}\mathrm{O(\alpha,\gamma)}^{22}\mathrm{Ne}$) will take place during the
merger when accretion results in high temperatures, and C and O may be
dredged up from the core.  A large amount of $^{18}\mathrm{O}$ will be
created only if this process is transient and is not allowed to proceed to
completion.  The available $^{14}\mathrm{N}$ is a result of CNO cycling in
the progenitor star, and the amount depends on the initial metallicity of
that star.  Hence the maximum amount of $^{18}\mathrm{O}$ formed cannot
exceed the initial abundance of $^{14}\mathrm{N}$, unless additional
$^{14}\mathrm{N}$ can be produced.
Our objective is to investigate whether the merger of a He WD and a CO WD
with a combined mass of $0.9M_\odot$ \citep[similar to RCB star
masses;][]{saio08} can lead to conditions suitable for producing oxygen isotopic ratios observed in RCB stars.

Close WD binary systems may be the progenitors for type Ia supernovae
\citep{iben84}, and such systems have therefore attracted much interest both
from theoretical and observational points of view.  Using an earlier
version of the hydrodynamics code used in this work, \citet{motl07} and
\citet{dsouza06} studied the stability of the mass transfer in close WD binary
systems.  

The fate of a close WD binary system depends on the mass ratio of the two
WDs.  Neglecting the angular momentum in the spin of the binary components
and allowing the angular momentum contained in the mass transfer stream to
be returned to the orbit, \citet{paczynski67} found that mass ratios below
$2/3$ are stable.  However, if the mass transfer stream directly strikes the
accretor instead of orbiting around it to form an accretion disk, this
stability limit may be reduced significantly.  Recent simulations by
Marcello et al.  (private comm.) indicate that even a mass ratio of 0.4 may
be unstable and lead to a merger.
\citet{brown11} have recently reported observations of the close WD binary
system, SDSS J065133.33+284423.3, which consists of a $0.25
M_\odot$ He WD and a $0.55 M_\odot$ CO WD, and is predicted to start mass
transfer in about 900,000 years.  If these two WDs merge, what will the
resulting object look like?

Other groups have employed smooth particle hydrodynamics (SPH) simulations
to study WD mergers, for instance \citet{benz90}, \citet{yoon07},
and \citet{raskin11}. In \citet{motl12}, the results from grid based
hydrodynamics simulations are compared to SPH simulations of
WD mergers, and it is found that the two methods produce results in
excellent agreement. Very recently, \citet{longland11} studied the
nucleosynthesis as the result of the merger of a $0.8 M_\odot$ CO WD and a
$0.4 M_\odot$ He WD.  The merger simulation was performed with an SPH
simulation code \citep{loren-aguilar09}.  They find that if only the outer
part of the envelope ($0.014<R/R_\odot<0.05$) is convective, the $^{16}\mathrm{O}$ to $^{18}\mathrm{O}$
ratio is 19, which is in the range measured for RCB stars
\citep{clayton07}.  On
the contrary, if the entire envelope is convective, the ratio is 370.

\citet{jeffery11} investigated the surface elements resulting from a merger
of a CO with a He WD, based on 1-D stellar evolution models and parametric nucleosynthesis analysis.  They considered two situations, a cold (no
nucleosynthesis) merger, and a hot merger (with nucleosynthesis).  In both
cases, they find surface abundances of C, N, and O that can be
made to match the observed RCB star surface abundances.  S and Si however,
do not match\footnote{\citet{asplund00} sugests that a possible solution is
condensation of dust which removes some gas phase abundance.}. In the hot merger scenario, the most promising location for nucleosynthesis to take place is in a hot
and dense region just on the outside of the original accretor, as for
instance seen in the simulations by \citet{yoon07}, \citet{loren-aguilar09},
or \citet{raskin11}.  This region forms as accreting matter from the donor
impacts the accretor.

  In this paper, we investigate whether the unusual abundances measured in
RCB stars can be produced in a WD merger, by first performing hydrodynamic
simulations of the merger of two WDs \citep[using a a modified version of
the 3-dimensional hydrodynamic code called Flower, see][]{motl02}.  In
section 2 of this paper, we present the methodology of our work.  Here the
details of the hydrodynamic code and the nucleosynthesis code along with 
their initial conditions are given.  In section 3, the
results of the hydrodynamic simulations and their corresponding
nucleosynthesis calculations are presented.  Finally, in section 4 we
compare our results with the results of other authors, and in section 5 we 
discuss our results along with future
directions for the work in this paper.

%
%
\section{Methods}


 In the hydrodynamic simulations, the fluid is modeled as a zero-temperature 
Fermi gas plus an ideal gas. 
The total pressure, $P$, is given by the sum of the ideal gas
pressure, $P_{\rm gas}$, and the degeneracy
pressure \citep{chandrasekhar39}: 
\begin{equation}
P_{\rm deg}=A\big[x(2x^2-3)(x^2+1)^{1/2}+3sinh^{-1}x\big],
\end{equation}
where $x=(\rho/B)^{1/3}$, and the constants A and B are given as:
\begin{eqnarray}
A&=\frac{\pi m_e^4 c^5}{3h^3}=6.00228\times10^{22} {\rm ~dynes~cm^{-2}} \\
B&=\frac{8\pi m_p}{3}\bigg(\frac{m_e c}{h}\bigg)^3=9.81011\times10^5 {\rm ~g~
cm^{-3}},
\end{eqnarray}
$m_e$ is the electron mass, $m_p$ is the proton mass, $h$ is Planck's
constant, and $c$ is the speed of light.
Similarly, the internal energy density of
the gas is the sum of the ideal gas internal energy density, $E_{\rm gas}$, 
and the internal energy density of the degenerate electron gas, 
$E_{\rm deg}$, \citep{benz90}:
\begin{equation}
E_{\rm deg}=
A\big[8x^3\big((x^2+1)^{1/2}-1\big)-\big(x(2x^2-3)(x^2+1)^{1/2}+3sinh^{-1}x\big)\big].
\end{equation}
The kinetic energy density of the gas is given by 
\begin{equation}
E_{\rm k}=\frac{1}{2}\rho v^2,
\end{equation}
where $v$ is the velocity of the fluid and $\rho$ is the density. The total
energy density, $E$, is the sum of these terms:
\begin{equation}
E=E_{\rm deg}+E_{\rm gas}+E_{\rm k}.
\end{equation}

The hydrodynamics code uses a cylindrical grid, with equal spacing between
the grid cells in the radial and the vertical directions.  We have run 5
simulations with the same total mass and different values of the mass ratio
(q) of donor to accretor mass of the two WDs.  The simulations with $q=0.7$,
$0.9$, and $0.99$ had 226 radial zones, 146 vertical zones and 256 azimuthal
zones, while the $q=0.5$ and $q=0.6$ simulations had 194 radial zones, 130
vertical zones, and 256 azimuthal zones.  The outer boundaries are
configured such that mass that reaches this boundary cannot flow back onto
the grid.  The outer radial boundary is about 1.5 times the size of the
outer edge of the donor WD (the larger star).  Likewise, the vertical
boundaries are about 2 times the size of the donor WD.  We allow the
hydrodynamics simulation to run sufficiently long after the two WDs have
merged that a steady-state-like configuration is reached.  The time to reach
this configuration depends, in part, on how much angular momentum is
artificially removed from the system.

Initially, the temperature is zero everywhere, meaning that $P_{\rm gas}$ and
$E_{\rm gas}$ are zero.
The initial $E$ can therefore be calculated from $\vec{v}$ 
and $\rho$ 
directly. The total energy density is evolved using the hydrodynamics equations,
from which $E_{\rm gas}$ at a subsequent time step can be found:
\begin{equation}
E_{\rm gas}= E-E_{\rm deg}-E_{\rm k}
\end{equation}
since both $\rho$ and $\vec{v}$ (needed to calculate $E_{\rm deg}$ and
$E_{\rm k}$) are also advanced using the hydrodynamics equations.
Knowing $E_{\rm gas}$ we can extract the temperature:
\begin{equation}
T=\frac{E_{\rm gas}}{\rho c_v},
\label{eq:T}
\end{equation}
where $c_v$ is the specific heat capacity at constant volume \citep{segretain97}
given by:
\begin{equation}
c_v=\frac{(<Z>+1)k_B}{<A>m_H(\gamma-1)}=1.24\times10^8{\rm ergs~g^{-1}~K^{-1}}
~\frac{(<Z>+1)}{<A>}=6.2\times10^7 {\rm ergs~g^{-1}~K^{-1}}
\label{eq:cv}
\end{equation}
where we have assumed that $(<Z>+1)/<A>=0.5$, with $<Z>$ and $<A>$ being the
average charge and mass for a fully ionized gas. Due to limitations in our
numerical approach, we will
assume an equal mix of C and O when calculating the temperatures. This is 
approximately correct for a CO mixture, and is a bit overestimated when He 
is present. However, when using the temperature for nucleosynthesis
calculations, we will use a corrected temperature taking helium and other
elements into account.

Using a self consistent field code developed by \citet{even09}, similar to
that developed by \citet{hachisu86a,hachisu86b}, a configuration of two
synchronously rotating WDs is constructed.  These data are used to
initialize the hydrodynamics simulations.  To speed up the merger process
(and in order to save CPU hours), angular momentum is artificially removed
from the system at a rate of $1\%$ per orbit for several orbits \citep[the
exact number of orbits depends on the simulation and does not seem to change
the outcome of the merger, see][]{motl12}.  This leads to mass transfer from
the donor star to the accretor, and, for the mass ratios that we
investigate, the stars end up dynamically merging.

 The results of the hydrodynamics simulations are inspected for locations
suitable for forming $^{18}\mathrm{O}$.  Using those conditions,
nucleosynthesis simulations are run using the post-processing network code
(PPN) from the NuGrid project \citep{herwig08}.
We use the single-zone frame (SPPN) of the NuGrid project
\citep{herwig08} to estimate the nucleosynthesis conditions
in the merger simulations as post-processing.  A nuclear network kernel, containing all the
nuclear reactions, reaction rates and a solver package, evolves the nuclear
network over each time step.  The input parameters required for the PPN code
are $T$, $\rho$, the initial abundances of nuclei, the time period over which
the network has to be calculated, and the time step for each calculation.  We
use the solar metallicity NuGrid RGB and AGB models (Set 1.2, Pignatari et
al.  in prep) that were calculated with the MESA stellar evolution code
\citep{paxton11} to derive the initial abundances of the shell of fire (SOF;
see section~\ref{resultsection}), where most of the nucleosynthesis takes
place, from the He-WD and CO-WD components.




We first look for the locations in the simulations that are conducive to
producing a high amount of $^{18}\mathrm{O}$, in order to obtain the
extremely low value of $^{16}\mathrm{O}$/ $^{18}\mathrm{O}$ observed in
RCBs.  The T, $\rho$, and the nuclear abundances of those locations are fed
as inputs to the nucleosynthesis code, which is then run over a suitable
period of time.  The evolution of various nuclear species relevant to this
ratio at the constant T, $\rho$ conditions chosen is studied.  The value of
the $^{16}\mathrm{O}$/ $^{18}\mathrm{O}$ ratio at each time step is then
compared to the observed value.

%
%
\section{Results}  
\label{resultsection}

\subsection{Hydrodynamics simulations}
\label{hydro_section}

\begin{figure}
\includegraphics{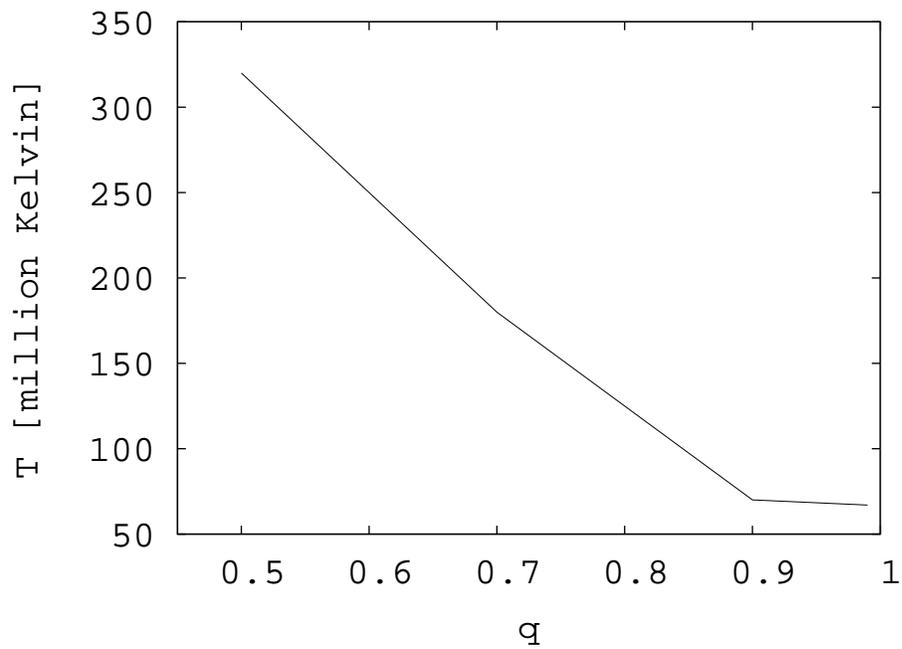}
\caption{The maximum temperature reached in the SOF (assuming C and
O only) that lasts for a considerable time plotted
versus the initial mass ratio $q$. For the high-$q$ simulations with no SOF post
merger, the temperatures are from the high density region surrounding the
core in the equatorial plane.}
\label{tvsq}
\end{figure}

When the initially cold donor material falls onto the cold accretor it is
heated through shocks or adiabatic compression.  This leads to a very hot
and dense region surrounding the accretor, the SOF (hence we labeled it the
``Shell of Fire'').  Such a SOF is a common feature in simulations of this
kind \citep[see for instance][]{yoon07, loren-aguilar09, raskin11}. 
However, only simulations with $q\lesssim0.7$ show a SOF around the merged
core.  During the simulations, the peak temperature found in regions with
high density\footnote{Our numerical approach can lead to artificial high
temperatures in low density regions in our simulations.  Since $t_{\rm
nuc}\sim 1/(\rho <\sigma_{\rm v}>)$ ($\sigma_{\rm v}$ being the nuclear
cross section) these hot, low density regions will not have significant
nuclear production.} strongly depends on the initial mass ratio (see
Fig.~\ref{tvsq}), with higher values of $q$ leading to lower temperatures
\citep[in disagreement with results in][]{dan12}.  In simulations with 
higher $q$ 
(and the same total mass), the material falling onto the accretor descends
into a shallower potential well.  Therefore, its kinetic energy is lower
when it impacts the accretor, leading to lower temperatures.  Typical
maximum temperatures in high density regions ($\rho\sim10^5{\rm g/cm^3}$) in
the high $q$ simulations are less than $2\times10^8$ K (assuming C and O
only) and last only for a short period of time comparable to an initial
orbital period of the system ($<100$ s).  The difference to
\citet{dan12} might be related to the fact that we do not take the nuclear
energy output into account. However, we note that our result is in agreement with
\citet{loren-aguilar09} who found that $T_{\rm peak}=6.5\times10^8$ K for
$q=0.5$ and $M_{\rm tot}=1.2 M_\odot$, while $T_{\rm peak}=6.3\times10^8$ K for
$q=1$ and $M_{\rm tot}=1.2 M_\odot$ and they too have nucleosynthesis in
their simulations. Another possible explanation for the difference with the
\citet{dan12} result is that they report the maximum temperature for minimum
$(\tau_{\rm nuc}/\tau_{\rm dyn})(T)$ ($\tau_{\rm nuc}$ and $\tau_{\rm
dyn}$ are the thermonuclear and dynamic time scales respectively), instead
of the peak temperature.

%
%


\begin{figure}
\includegraphics[width=\textwidth]{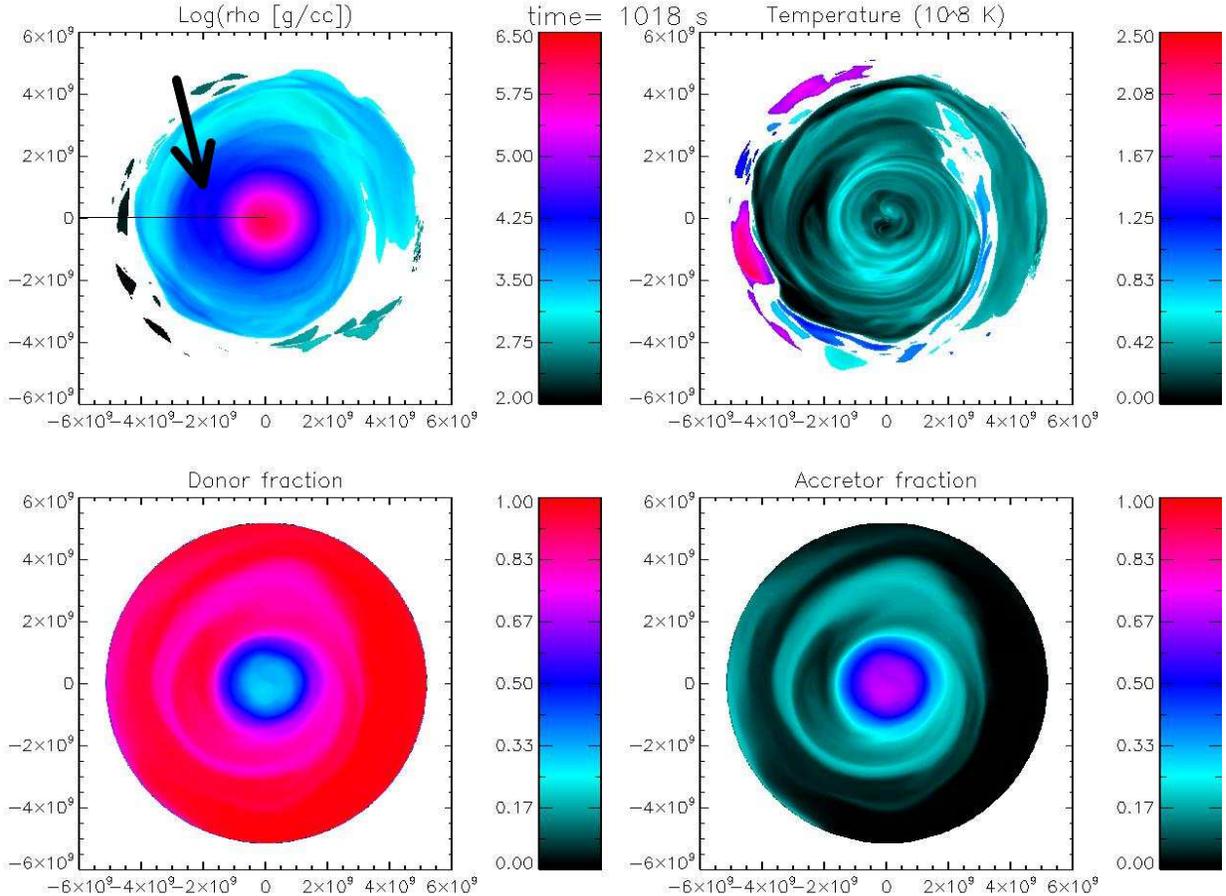}
\caption{The $q=0.9$ simulation after the two stars have merged showing
log density (upper left), temperature (upper right),  mass
fraction of donor material (lower left), and  mass fraction of accretor
material (lower right) in the
equatorial plane. The horizontal line in the density plot illustrates where
the plot in Fig.~\ref{q0.9vertfigure} is made. Much of the
donor material is violently digging into the accretor core (density larger
than $10^{5.2} {\rm g/cm^3}$) during the merger. 
In the process, accretor
material is being dredged up, mixing with the
incoming donor material mostly in a layer just outside the core. The central
region with a density larger than about $10^4 {\rm g/cm^3}$ can be seen to be
somewhat asymmetric. A colder, donor rich ``blob'' of material sits to the left
of the central high density core (around $2\times10^9$ cm from the center),
indicated by the arrow in the density plot.
This blob is found to be more prominent in the $q=0.7$ simulation.}
\label{q0.9figure}
\end{figure}

\begin{figure}
\includegraphics[width=\textwidth]{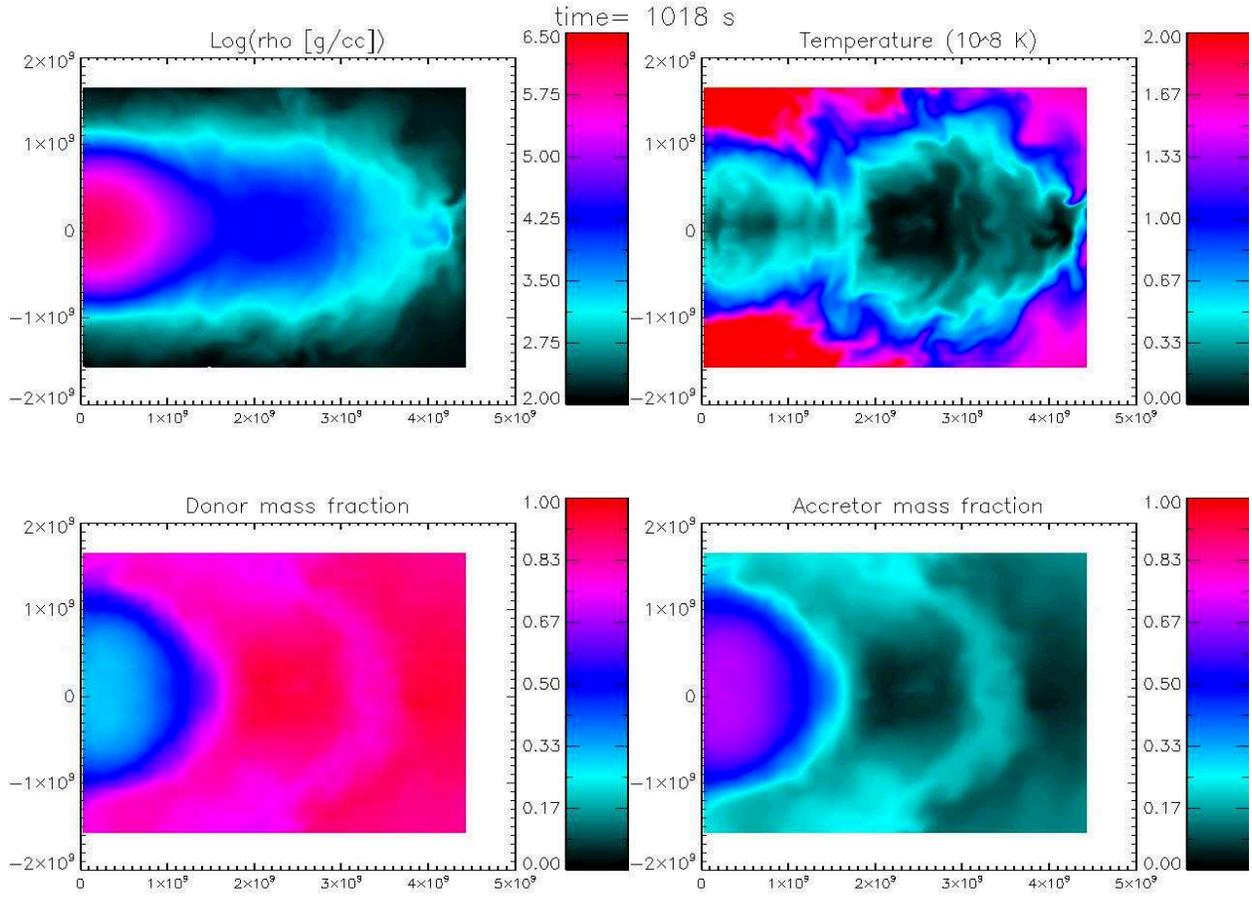}
\caption{The $q=0.9$ simulation after the two stars have merged, showing
log density (upper left), temperature (upper right),  mass
fraction of donor material (lower left), and  mass fraction of accretor material (lower right) in the
r-z plane from the center of the grid outwards, intersecting the blob 
(along the line in the density plot 
in Fig.~\ref{q0.9figure}). Little mass is being forced out in 
the vertical direction, explaining the very
low densities far from the equator.}
\label{q0.9vertfigure}
\end{figure}

We now discuss the details of the high and low $q$ simulations, choosing
$q=0.9$ and $q=0.7$ as our representative cases\footnote{We note that even
though the $q=0.5$ and $0.6$ simulations have a slightly different resolution,
this does not appear to affect the results.}.  The merger in the high $q$
simulations (Figs.~\ref{q0.9figure} and \ref{q0.9vertfigure}) is extremely
violent, and the accretor core is severely distorted by the incoming
accretion stream.  A thin layer just outside of the combined core contains a
mixture of donor and accretor material. Assuming a
composition of carbon and oxygen only, 
the temperature in the high density regions does not reach $10^8$ K
other than in a few transient areas during and after the merger.

\begin{figure}
\includegraphics[width=\textwidth]{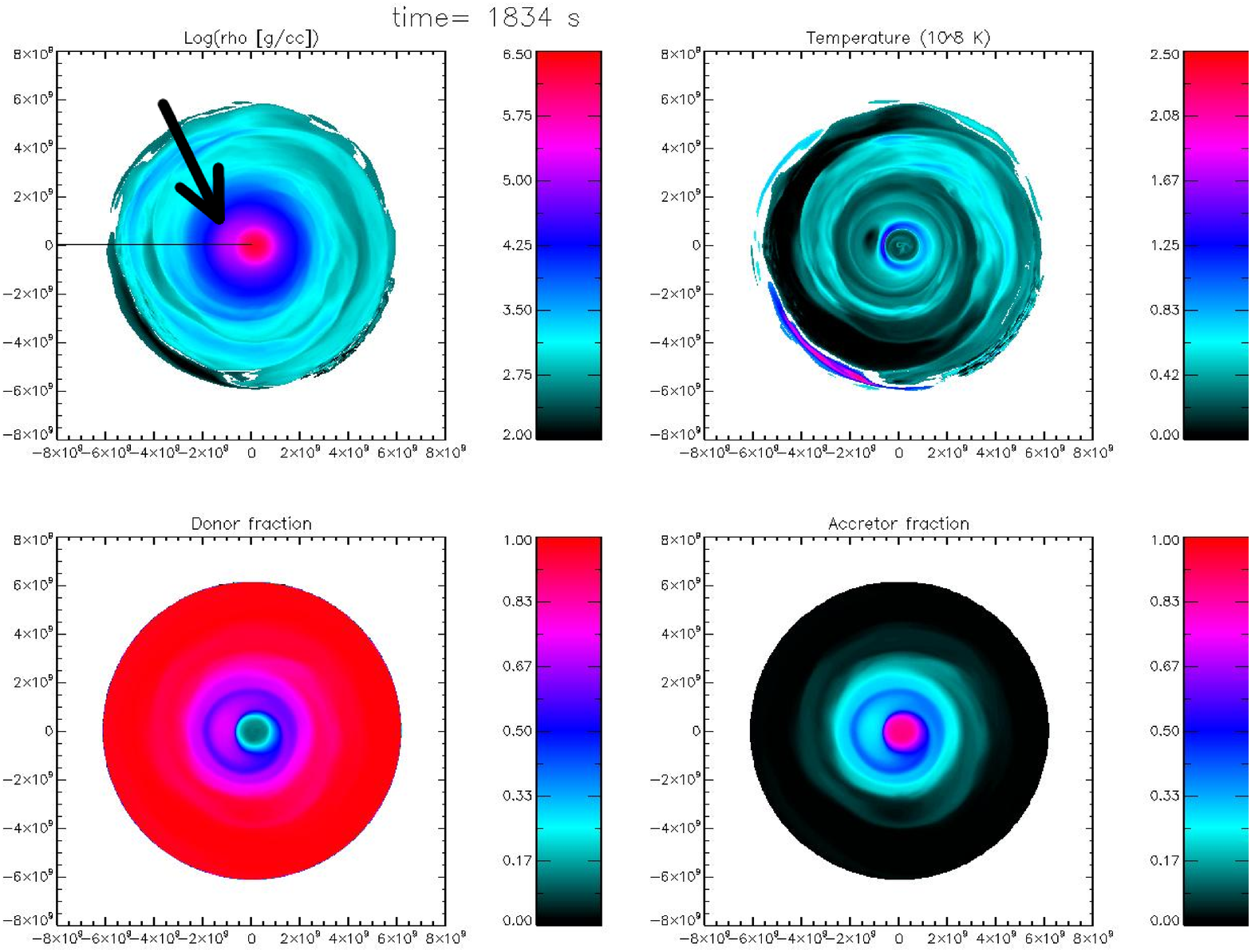}
\caption{The $q=0.7$ simulation after the two stars have merged, showing
log density (upper left), temperature (upper right), mass
fraction of donor material (lower left), and mass fraction of accretor
material (lower right) in the
equatorial (r-$\phi$) plane. The horizontal line in the density plot
illustrates where the plot in Fig.~\ref{q0.7vertfigure} is made. Compared to 
the higher-$q$ simulations, there are noticeable differences in this simulation.
A hot SOF forms around the merged core with temperatures up to 
$1.5\times10^8$ K (assuming C and O only; from Fig.~\ref{q0.7vertfigure} we can see that
this is actually a SOF). The merger
is also less violent, in that the accretor core is not distorted as much as
in the higher $q$ cases. However, much accretor material is being
dredged up during the merger. The core is seen to be
asymmetric. A cold, donor-rich blob sits to the left of the
central, highest density core, indicated by the arrow in the density plot.}
\label{q0.7figure}
\end{figure}

\begin{figure}
\includegraphics[width=\textwidth]{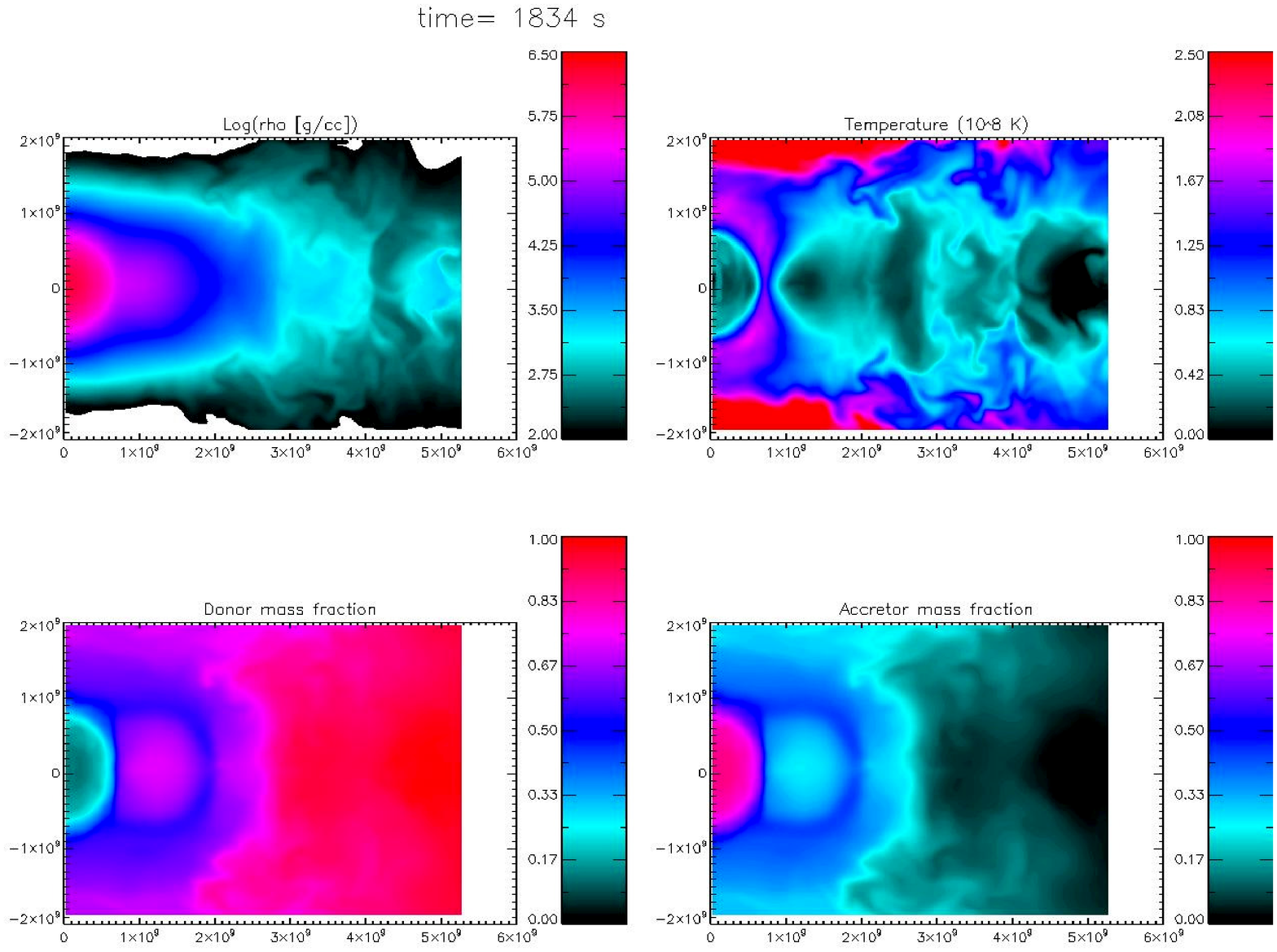}
\caption{The $q=0.7$ simulation after the two stars have merged, showing
log density (upper left), temperature (upper right), mass
fraction of donor material (lower left), and  mass fraction of accretor
material (lower right) in the 
r-z plane, intersecting the blob (along the line in the density
plot in Fig.~\ref{q0.7figure}). 
A hot SOF develops around the merged core, with temperatures up 
to about $1.5\times10^8$ K (assuming C and O only). Even though the core is not distorted as much as
in the higher $q$ simulations, we notice that much accretor and donor 
material is pushed up vertically from the core. The donor rich blob is 
clearly visible
in the mass fraction plots of donor and accretor material, between about $8\times10^8$ cm
and $2\times10^9$ cm. It can also be seen in the density plot.}
\label{q0.7vertfigure}
\end{figure}

Lower $q$ simulations (see Fig.~\ref{q0.7figure} and
Fig.~\ref{q0.7vertfigure}) show a much less violent merger.  Even though the
accretor core does not get significantly distorted, a large amount of
accretor material is being dredged up and mixes with the incoming donor
material.  Indeed, the donor star is tidally disrupted before the cores
merge.  This helps to preserve the SOF in these cases, as the donor material
is added more gently on top of the accretor, instead of
falling through the SOF to mix in with the core as in the high $q$
simulations.  On the other hand, in the high $q$ simulations the two cores
merge destroying the SOF in the process to form the newly merged
core\footnote{Movies of all the simulations showing density, temperature,
and mass ratios in the equatorial plane can be found here:\\ {\tt
http://phys.lsu.edu/$\sim$astroshare/WD/index.html}}, by mixing the hot
pre-merger SOF material with cold donor and accretor core material.
 
An asymmetric feature in the merged object becomes very clear in the $q=0.7$
simulation; looking at the density plot in Fig.~\ref{q0.7figure}, a lower
density blob can be seen extending from the core in the negative ``x''
direction.  This blob has a very low temperature, and to be donor
rich/accretor poor (as such it can be thought of as some of the last of the
donor material, accreted but never heated).  It is
encapsulated by accretor rich material.  This is also clearly visible
in Fig.~\ref{q0.7vertfigure}, which shows a slice taken directly through the
blob.  A blob is also present in the other simulations (both high and
low $q$), although it appears most prominent in the $q=0.7$ configuration. The
SOF has formed between
this blob and the merged core, with sustained
temperatures of $\sim 1.5\times10^8$ K or more (assuming C and O only) lasting at
least for the duration of the $q=0.7$ simulation (similar features are found in
the $q=0.5$ and $0.6$ simulations).  In the $q=0.6$ simulation, we
find sustained temperatures in the SOF of about $2.5\times10^8$ K, while in the
$q=0.5$ simulation the sustained temperatures are about $3\times10^8$ K. 
We find the SOF (in all low $q$ cases) to be located just outside the
merged core ($r\sim10^9$ cm), with a thickness of about $1-2\times10^8$ cm
(Table~\ref{sofconditions}).  We assume the
core of the merged object is where $\rho>10^{5.2}{\rm g/cm^3}$ and $T<10^8
{\rm K}$, while the SOF is defined as being $\rho>10^{4.25}{\rm g/cm^3}$
and $T>10^8 {\rm K}$.  The core density value was chosen so that it
extends out to the SOF.  As we will see in the next section, the low $q$
simulations turn out to be the cases relevant to the nucleosynthesis of
$^{18}\mathrm{O}$.  Table~\ref{lowqdetailstable} lists the details of
the cases and their SOFs.



\begin{table}
\caption{Summary of the conditions found in the SOF after the merger. In particular,
$q$ is the mass ratio 
used in the hydrodynamic simulations, $T_{\rm SOF}$ and $\rho_{\rm SOF}$ are representative values of density and temperature in the SOF, $f_{\rm sof, acc}$ represents the fraction of the SOF made of the accretor material and  $dR_{\rm SOF}$ refers to the (approximate) width of the SOF. }
\begin{tabular}{ccccc}
q & $T_{\rm sof}$ ($10^6$ K) & $\rho_{\rm sof} ({\rm g/cm^3})$ &
$f_{\rm sof, acc}$ & $dR_{SOF} (10^9)$ cm\\ \hline 
0.5 & 300 & $10^{4.5} $ & 0.67 & 0.2 \\
0.6 & 250 & $10^{4.5} $ & 0.54 & 0.15 \\
0.7 & 150 & $10^{4.7} $ & 0.50 & 0.2 \\
\hline
\end{tabular}
\label{sofconditions}
\end{table}


\begin{table}
\caption{For the simulations, we list the mass of the accretor WD
(CO; $M_{CO}$), the mass of the donor WD (He WD in lower $q$ simulations; 
$M_{He}$), the mass of the SOF
($M_{\rm SOF}$), the mass that is present outside the merged core and the
SOF ($M_{\rm out}$), the time from the beginning of the simulation to the end of  the
merger ($t_{merge}$), the total time for which the simulation was run $t_{end}$, and the time for which the simulation ran after the merger ($\delta \rm t$ = $t_{end} -t_{merge}$). }
\begin{tabular}{cccccccc}
$q$ & $M_{CO}/M_{\odot}$ &$M_{He}/M_{\odot}$& $M_{\rm SOF}/M_{\odot}$ &$M_{\rm
out}/M_{\odot}$&$t_{merge}(s)$& $t_{end}(s)$ &$\delta \rm t(s)$ \\ \hline 
0.5 & 0.6  & 0.30 & 0.12 &0.24 &2070 & 2542 & 472 \\
0.6 & 0.56 & 0.34 & 0.13 &0.26 &1150 & 1500 & 350 \\
0.7 & 0.53 & 0.37 & 0.10 &0.30 &1200 & 1970 & 570  \\
0.9 & 0.47 & 0.43 & none & 0.34 & 667 & 1084 & 417 \\
0.99 & 0.45 & 0.45 & none & 0.34 & 698 & 1137 & 439 \\
\hline
\end{tabular}
\label{lowqdetailstable}
\end{table}

\begin{figure}
\includegraphics[width=0.8\textwidth]{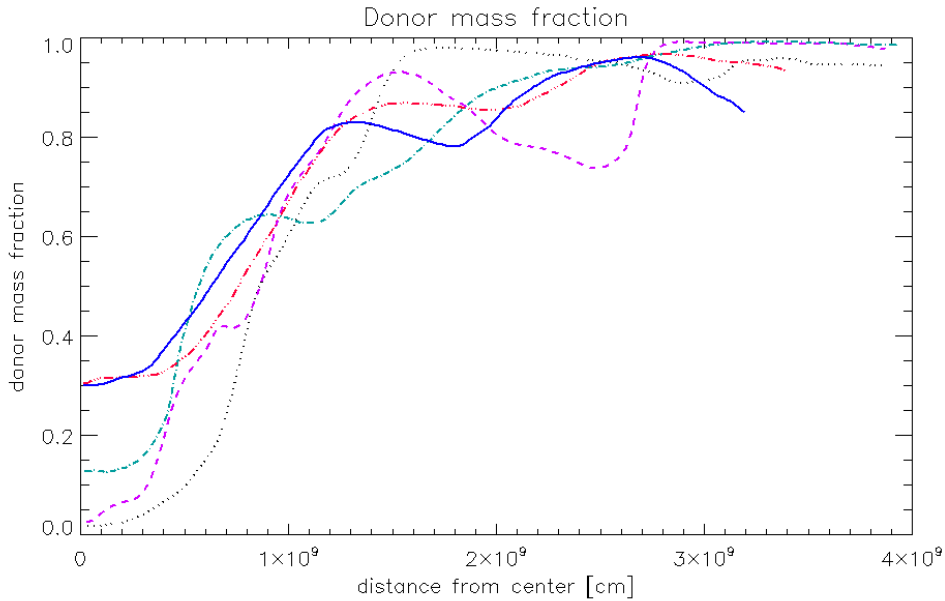}
\caption{ The mass fraction of the donor material as a function of the distance from the
center of the merged object at an arbitrary time after merger. The curves are for 
$q=0.5$ (dotted; black), $q=0.6$ (dashed; purple), $q=0.7$
(dash-dotted; light blue), $q=0.9$ (solid; blue), and $q=0.99$
(dash-triple-dotted; red). The core is about $10^9$ cm, 
surrounded by an SOF
for the $q=0.5$, $0.6$, and $0.7$ cases. The mass fraction of the
accretor material for each
of these cases may be obtained as 1- mass fraction of the donor material.}
\label{donormassfractionplotfigure}
\end{figure}

\begin{table}
\caption{For each of the low-$q$ simulations, the fraction of mass
($ f_\mathrm{acc-dredged}$) dredged up from the accretor, the accretor mass in the SOF
($m_{\rm
acc_SOF}/M_{\odot}$) in solar masses  and the accretor mass
that is outside the SOF ($m_{\rm acc_out}/M_{\odot}$).}
\begin{tabular}{cccc}
$q$ & $f_{\rm acc-dredged}$ & $m_{\rm acc-SOF}/M_{\odot}$  & $m_{\rm
acc_out}/M_{\odot}$  \\ \hline 
0.5 & 0.17 & 0.08 & 0.02\\
0.6 & 0.18 & 0.07 & 0.03 \\
0.7 & 0.15 & 0.05 & 0.03  \\
\hline
\end{tabular}
\label{dredgeuptable}
\end{table}

In Fig.~\ref{donormassfractionplotfigure}, we plot the donor
mass fraction as a function of radius in the equatorial plane, where the 
 mass fraction of the accretor material is simply one minus the mass fraction of the donor
material.  In all
simulations (including the low $q$ cases), we find that a significant amount of
accretor material is being dredged up and mixed with the donor material
outside of the merged core.  The dredged up accretor material leads to a layer
just outside the core that is heavily enriched with accretor material.  
 Outside the merged core and the SOF, the mass of accretor material is
nearly the same in all the low $q$ simulations 
(Table~\ref{dredgeuptable}).  From Figs.~\ref{q0.9figure} and
\ref{q0.9vertfigure} 
we see that the high $q$ simulations have considerable mixing in their
cores due to the very violent mergers forming them.  The low $q$ simulations do
not experience such violent mixing.


\subsection{R Coronae Borealis progenitor systems}
\label{progenitor_system}

In order to perform nucleosynthesis calculations, we use the temperature and
density conditions of the SOFs.  Post merger, the SOFs are seen as a feature 
solely of the low $q$ cases, with temperatures ranging from
$1.5\times10^8$ to
$3\times10^8$ K (Fig.~\ref{tvsq}) with densities between ${\rho} = 10^{4.25} -
10^{5.2} {\rm g/cm^3}$ (Table ~\ref{sofconditions}).  Such conditions
make the SOF a favorable site for the production of
$^{18}\mathrm{O}$ and are used as inputs to the
nucleosynthesis code.


\begin{figure}
\includegraphics{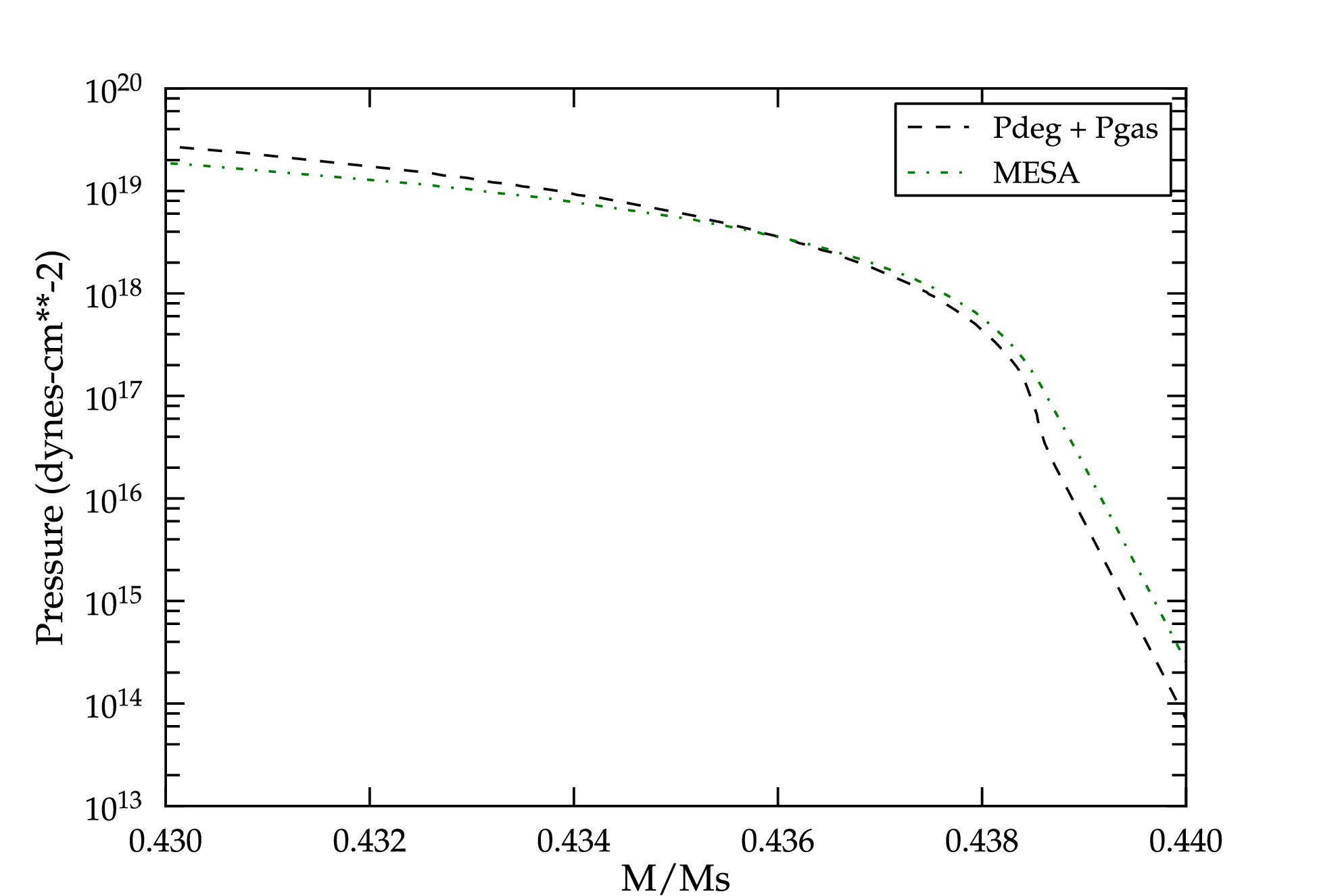}
\caption{Pressure profile of a model in the 11th thermal pulse on the AGB of an  initial mass of 2 $M_{\odot}$, compared against the pressure calculated using the EOS used in the hydrodynamic simulations. The region shown here is typically where the SOF is located during the course of the merger simulations.}
\label{eos}
\end{figure}

Since the SOF is the region where we focus our analysis, we do a simple
validation test of the EOS used in the hydrodynamic simulations.  To do
this,
we compare the pressure calculated using this EOS, for the density and
temperature of the H-free core of a MESA computed AGB model on its 11th
thermal pulse having an initial mass of 2 $M_{\odot}$, against the pressure
profile given by MESA for the same model.  Fig.~\ref{eos} shows the outer
portion of the CO WD where the SOF appears in the simulations.  We can
see the pressure of the CO WD model from MESA is very close to that obtained
from the EOS used in the hydrodynamic simulations, the pressure being slightly
lower within 0.436 $M_{\odot}$ and higher above it.  This gives a
reasonable confirmation on the choice of the EOS for the purpose of these
simulations.

For the nucleosynthesis calculations, the initial chemical abundances
in the SOF are required.  According to the hydrodynamic
simulations, the SOF has contributions from the He WD as well as material
dredged up from the CO WD.  The initial abundances of the WDs
are calculated using realistic 1D stellar evolution models computed with
MESA \citep{paxton11} and post processed with NuGrid codes.


\begin{figure}
\includegraphics[scale=0.65]{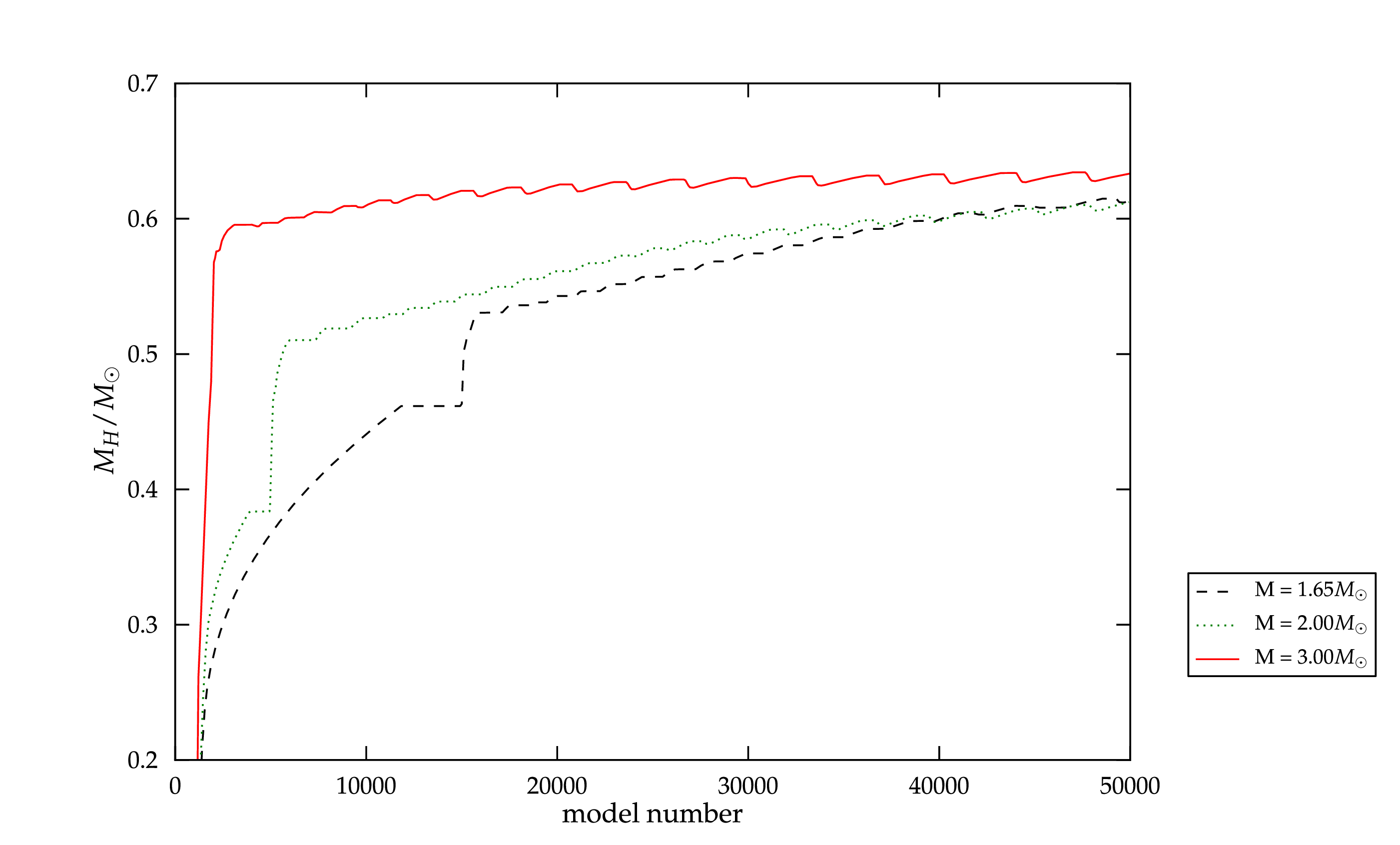}
\caption{The evolution of the hydrogen free core mass ($M_{H}/M_{\odot}$) 
during the AGB phase of the star, for three initial masses. The model numbers on the horizontal axis are related to time, with a higher model number being a later time.
 The AGB thermal pulse phase occurs over a much smaller time period than the previous evolution of the star. Hence, if age were used as the x axis parameter, the thermal pulse features would be compressed and not visible. Model numbers help in visualising the entire range of evolution of the star.}
\label{MH_mod}
\end{figure}
\begin{figure}

\includegraphics[scale=0.65]{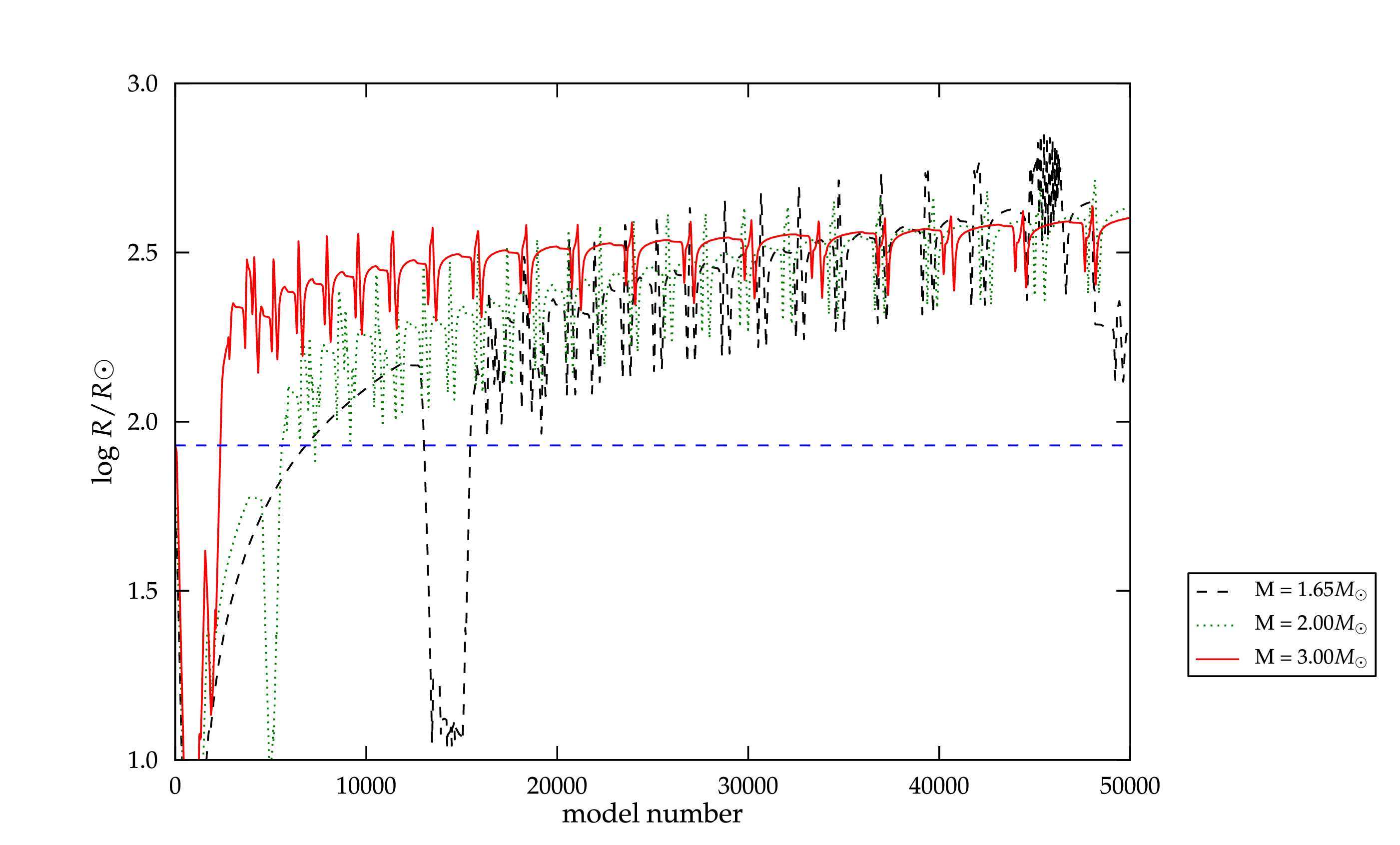}
\caption{The variation of the radius of the star ($\mathrm{log R/R_{\odot}}$) 
plotted against the model number, during the 
AGB phase of the star, for three initial masses.
 The blue dashed line is the radius of the progenitor RGB star of the He WD under consideration. A model number corresponds to every step in time taken by the code during the run for each initial mass; a higher model number corresponds to at a later point of
time.}
\label{R_mod}
\end{figure}

In order to determine the initial abundance contribution from the CO WD, the
evolutionary state of the AGB progenitor from the last common envelope (CE) phase has to be
determined.  During the likely binary progenitor
evolution that leads to the double degenerate merger considered here, 
one or more CE phases can occur
\citep{iben84}.  For the scenario that we consider, there are two CE
phases.  The first one occurs when the primary star overflows
its Roche lobe during its AGB phase, thus forming the CO WD.  For a star to
fill its Roche lobe, its radius must be greater than or equal to its Roche
lobe radius ($R_{L}$).  $R_{L}$ is the product of a function of the
mass ratio $E(q)$ between the primary and the secondary
components of the binary system and the separation ($a$) between them, given
by $R_{L}=E(q)a$ \citep{eggleton83}.  From observations of binary systems,
the separation between components can range between $3 - 10^{4} R_{\odot}$
\citep{hurley02}, which implies that the $R_{L}$ of a star in a binary
system, with a given initial mass ratio, can vary over 4 orders of magnitude
depending on the separation distance between the two components.


\begin{figure}
\includegraphics[width=\textwidth]{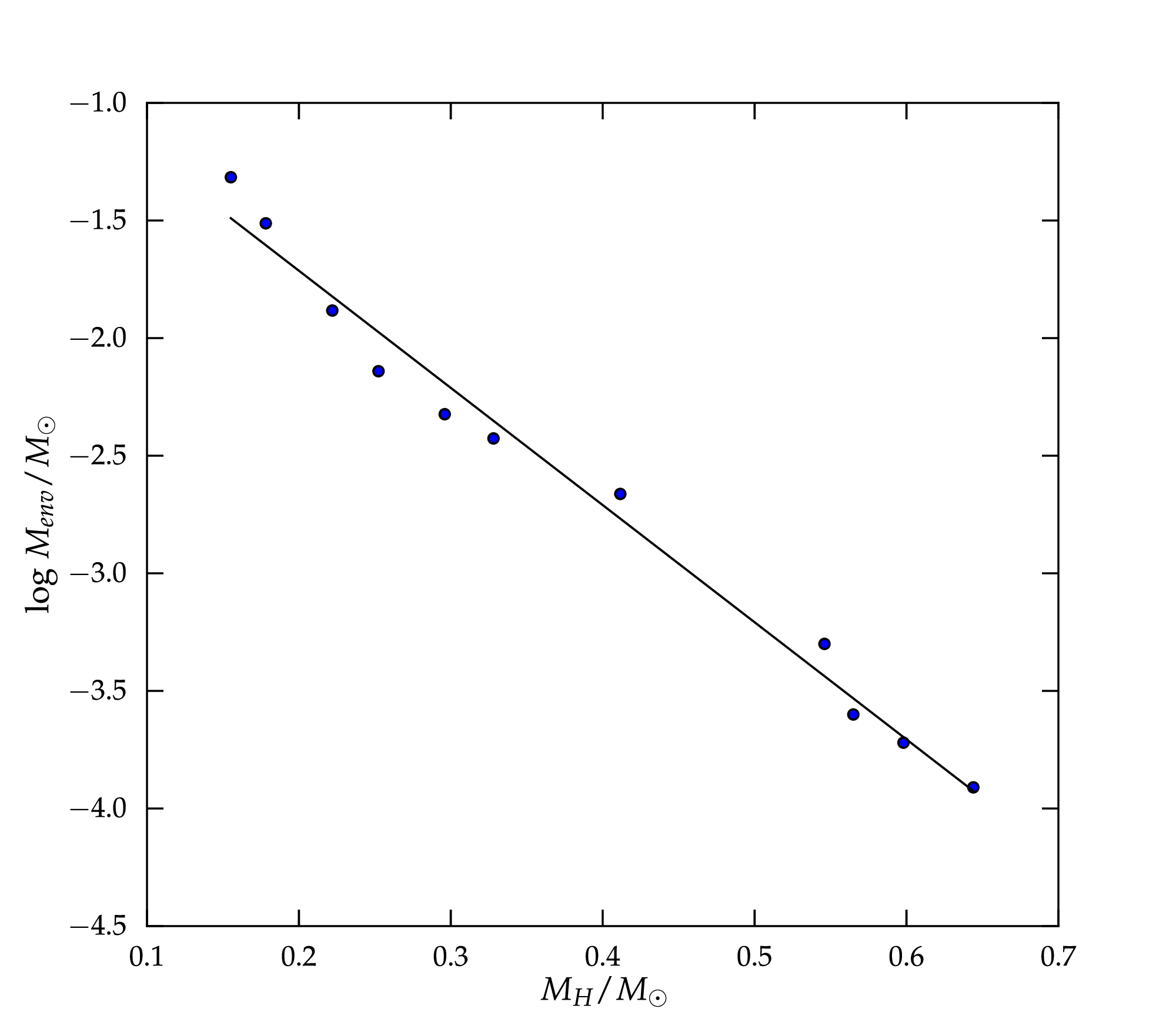}
\caption{Plot showing the anti-correlation between envelope mass (log $M_{env}/M_{\odot}$ and hydrogen-free core mass $M_{H}/M_{\odot}$. A best fit line using the least squares method is drawn through the blue data points.}
\label{menv_mh}
\end{figure}

From stellar evolution studies, the structure of a WD consists of a hydrogen
free core, $M_{H}$, surrounded by a thin envelope of unprocessed material
which as a
result is rich in hydrogen.  The envelope mass is typically between
0.05 to 1$\%$ of the mass of the WD and is anti-correlated with 
$M_{H}$ where the outer boundary 
is the radial co-ordinate at which the hydrogen abundance $X_{H}$
= 0.37 \citep{schon83}.  Using the data from the work of \citet{schon83} for
CO WDs and \citet{driebe98} for He WDs, a least squares fit is done between
the data points.  The best fit line thus constructed enables us to read off
the envelope mass, $M_{env}$, for a particular $M_{H}$ (between 0.1552 and
0.644 $M_{\odot}$) (Fig.  \ref{menv_mh}).  The analytic equation of this
line is $\log(M_{env}/M_{\odot})=-4.982M_{H}/M_{\odot} - 0.7171$ .

The hydrodynamic simulations of interest for nucleosynthesis, use a range of
masses of CO WDs ($M_{CO}$) between 0.53 and 0.6 $M_{\odot}$.  For the
purposes of the following explanation, we can take $M_{CO}= M_{H}$
since the envelope mass
($M_{env}$) of the CO WD is less than $0.1\%$ of its total mass.

 However, knowing the white dwarf mass does not imply knowledge of the
initial mass of its progenitor star.  Figure~\ref{MH_mod} shows the
evolution of $M_{H}$ (${\sim} M_{WD}$) (NuGrid Set1.2 models for solar
metallicity, Pignatari et al.  2012, in preparation) for a range of stellar
masses ($1.65, 2$ and $3 M_{\odot}$). It is evident that for a given value of
$M_{H}$ the star could have any of the three initial masses and can lie
anywhere between the early AGB phase and a late thermal pulsing (TP) phase. 
The higher the number of TPs that the star has undergone, the more enriched
it is in partial He-burning and s-process products \citep{herwig05}.  A
parameter that will help in solving this degeneracy between the initial and
the core mass of a star is the Roche lobe radius of the giant star that
first fills its Roche lobe in the binary system.  Fig.~\ref{R_mod} shows the
radius evolution of the AGB stellar model sequences for the same initial
masses.

Let us consider an He WD model from the cases plotted, whose mass is nearly
the same as the one in the $q=0.5$ simulation.  This model has mass
$M_{He}=0.3024 M_{\odot}$ from a star of initial mass of 1.65 $M_{\odot}$. 
 If we assume that the RGB progenitor of this He WD fills its Roche lobe
when it reaches this radius and that the binary system enters its second CE
phase immediately when it does so, then it is a good estimate that the
maximum separation distance between the two components is at the most, the
Roche lobe radius ($\mathrm{log~R_{L}/R_{\odot}}$) of the secondary, which
is 1.286 in this case.

From previous hydrodynamic simulation work done by \citet{demarco11},
we know that after the binary system has undergone its first CE event, the
separation between the components reduces by at least 4.5 times its initial
separation.  The minimum separation at the time of the first CE event is at
least $\mathrm{log~a/R_{\odot}}$ = 1.93.  We assume that the binary system
enters its first CE phase immediately when the AGB star fills its Roche
lobe.  Since ($\mathrm{log~a/R_{\odot}}$) = 1.93 is the minimum limit on the
Roche lobe radius of the AGB star, we investigate three cases during
different phases of the AGB star, when its radius exceeds this value.  These
are, during the early AGB phase (CO WD(1)), an early TP phase (CO WD(2)), and
a late TP phase (CO WD(3)) (after the star becomes carbon rich)
(Fig.~\ref{R_mod}).  It must be pointed out that the post CE WD abundance
profile is assumed to be that of the inner portion of the progenitor AGB
models considered here.  Henceforth the CO WD models are to be understood as
the progenitor AGB stars with mass equal to $M_{CO}= M_{H}$ + $M_{env}$.

Table ~\ref{CO_WD_mod} summarizes the relevant parameters of these CO WD models.  It must be noted that while the star is in the TP phase
the chosen model must be at the peak of the pulse, since the radius of the star is at its maximum during the peak of a given pulse. If the star has not been able to fill its Roche lobe during an earlier pulse peak, it cannot do so until it hits the next pulse peak.

From the hydrodynamic simulations it is seen that the He WD is totally disrupted and well mixed during the final merging phase.  Hence for the He WD the nuclear abundances are averaged over its entire mass, thus giving a uniform
composition for the He WD. From Table ~\ref{dredgeuptable}, the fraction of CO WD dredged up outside is less than $20\%$.  Table ~\ref{prog_abun} contains the isotopic abundances of the He WD and the cumulative abundances of significant elements in the outer $20\%$ of the CO WD models.

\begin{figure}
\includegraphics[width=\textwidth]{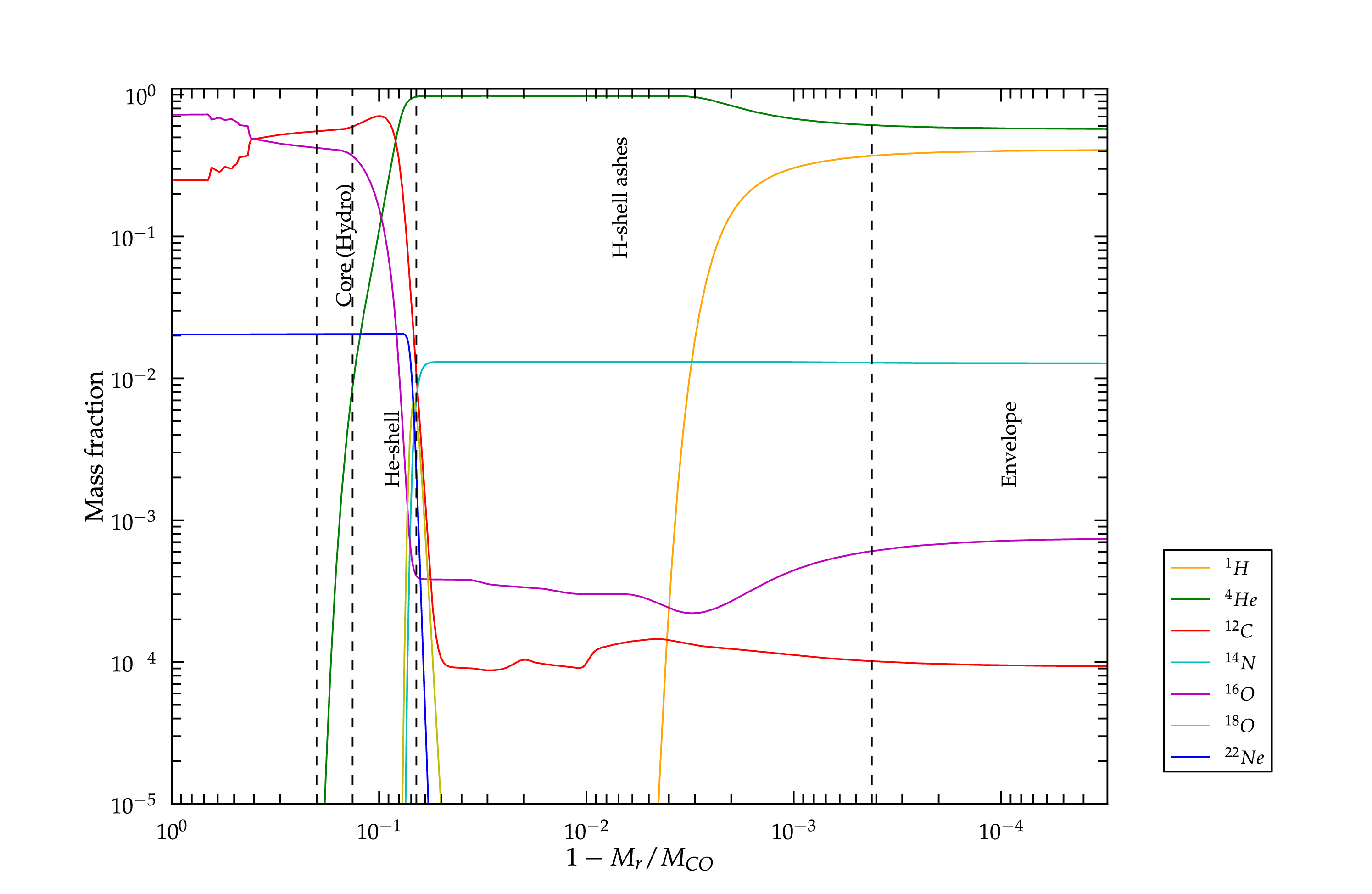}
\caption{Nuclear abundance profile of the CO WD (1) model. The x-axis is $(1-M_{r}/M_{CO})$ where $M_{r}$ is the radial mass co-ordinate and $M_{CO}$ = 0.58148. The zones of the CO WD are labelled and marked with dashed black lines. The lower limit on the zone marked
``Core (hydro)'' is placed at a depth enclosing $20\%$ of the $M_{CO}$ mass, which is the upper limit on the mass dredged up of the CO WD as seen in the hydrodynamic simulations.}
\label{E_agb}
\end{figure}

In order to achieve an extremely low $^{16}\mathrm{O}/^{18}\mathrm{O}$ ratio
such as that observed in RCB stars, we take the CO WD model which provides
the abundances most viable to help realize this.  The model which has the
highest amount of $^{14}\mathrm{N}$ and $^{18}\mathrm{O}$ and the least
amount of $^{16}\mathrm{O}$, amongst the three cases is selected.  This
model belongs to the early AGB phase of the 3 $M_{\odot}$ star ,
(Fig.~\ref{E_agb}) and has $M_{CO}$=0.58148 $M_{\odot}$ (CO WD(1), Table
~\ref{CO_WD_mod}) and an envelope of $2.5\times10^{-4} M_{\odot}$.  It must
be noted that since the progenitor of the CO WD model chosen is on the
E-AGB, it does not have any s-process element enhancement on it's surface. 
Hence although the choice of this CO WD model may lead to the reproduction
of the unique O isotopic ratio values of RCB stars, it may not reproduce the
s-process element enrichment found in them.

Thus we have used the CO WD and He WD models described above, to provide the
initial abundances of the SOF.  With the knowledge of the the fraction of
the SOF made of the accretor ($f_{\rm sof,acc}$) and the donor (1-$f_{\rm
sof, acc}$, Table~\ref{sofconditions}), the initial abundances of the SOF
are constructed (Table ~\ref{initial_abun}).  We assume that all the
material from the CO WD has been dredged up from its outer layers, before
the onset of hot nucleosynthesis in the SOF.

\begin{table}
\caption{The details of the CO WD models - initial mass ($M_{\star}$), model number, phase of evolution, mass of the CO WD ($M_{CO}/M_{\odot}$), mass of hydrogen free core ($M_{H}/M_{\odot}$), and $\mathrm{log~ R/R_{\odot}}$.}
\begin{tabular}{ccccccc}
Serial number & $M_{\star}$ &  model number & phase of evolution & $M_{H}/M_{\odot}$ & $M_{CO}/M_{\odot}$ & $\mathrm{log~R/R_{\odot}}$\\ \hline 
1              &3& 2364   & E-AGB    	         & 0.58123  &0.58148   & 1.97 \\
2	       &2& 12198  & 3$\mathrm{rd}$ TP    & 0.53334  &0.53376   & 2.35 \\
3	       &2& 49901  & 21$\mathrm{st}$ TP   & 0.61226  &0.612243   & 2.65 \\
\hline
\end{tabular}
\label{CO_WD_mod}
\end{table}


\begin{table}
\caption{ The averaged (approximate) abundances ($\%$) by mass, for dominant and relevant 
species for the evolution of the abundance of $^{18}\mathrm{O}$ in the outer part of the 
three CO WDs are presented in Table~\ref{CO_WD_mod} (we considered the most external zone, 
which corresponds to about $20\%$ of the total WD masses). The other nearly 300 species in 
the nuclear network of the code, are present in much smaller proportions.}
\begin{tabular}{ccccc}
Species           & He WD    & CO WD (1) & CO WD (2) & CO WD (3) \\ \hline 
$^{1}\mathrm{H}$  & 1.51      & 0.33    & 0.276     & 0.093 \\
$^{4}\mathrm{He}$ & 96.5     & 41.7    & 14.6      & 9.04 \\
$^{12}\mathrm{C}$ & 0.011   & 36.5   & 50.5     & 43.9\\
$^{14}\mathrm{N}$ & 1.3      & 0.43  & 0.061    & 0.155  \\
$^{16}\mathrm{O}$ & 0.074    & 19.0   & 32.0     & 43.8 \\
$^{18}\mathrm{O}$ & 2.86${\times}10^{-5}$ & 2.7${\times}10^{-3}$ & 2.71${\times}10^{-4}$ &1.11${\times}10^{-5}$   \\
$^{22}\mathrm{Ne}$ & 2.73${\times}10^{-4}$ & 1.34    & 1.95      & 2.18 \\
\hline
\end{tabular}
\label{prog_abun}
\end{table}


\begin{table}
\caption{ The averaged initial abundances ($\%$) for dominant and relevant species for the evolution of the abundance of $^{18}\mathrm{O}$ in the SOF of the $q=0.5$, $0.6$, and $0.7$ cases using the He WD and CO WD(1) models of Table~\ref{prog_abun}. The other nearly 300 species in the nuclear network of the code, are present in much smaller proportions.}

\begin{tabular}{cccc}
Species        &   SOF,0.5 & SOF,0.6  & SOF,0.7 \\ \hline 
$^{1}\mathrm{H}$ & 0.75    & 0.89     & 0.97      \\
$^{4}\mathrm{He}$ & 64.8    & 69.5    & 75.9 \\
$^{12}\mathrm{C}$ & 22.2    & 18.6     & 15.1 \\
$^{14}\mathrm{N}$ & 0.76   & 0.85     & 0.92 \\
$^{16}\mathrm{O}$ & 10.03    & 8.9     & 5.9 \\
$^{18}\mathrm{O}$ & 0.021  & 0.016   & 0.018 \\
$^{22}\mathrm{Ne}$ & 0.81   & 0.68    & 0.55 \\
\hline
\end{tabular}
\label{initial_abun}
\end{table}


\subsection{Nucleosynthesis}
\label{nucleosynthesis}

\begin{figure}
\includegraphics[width=\textwidth]{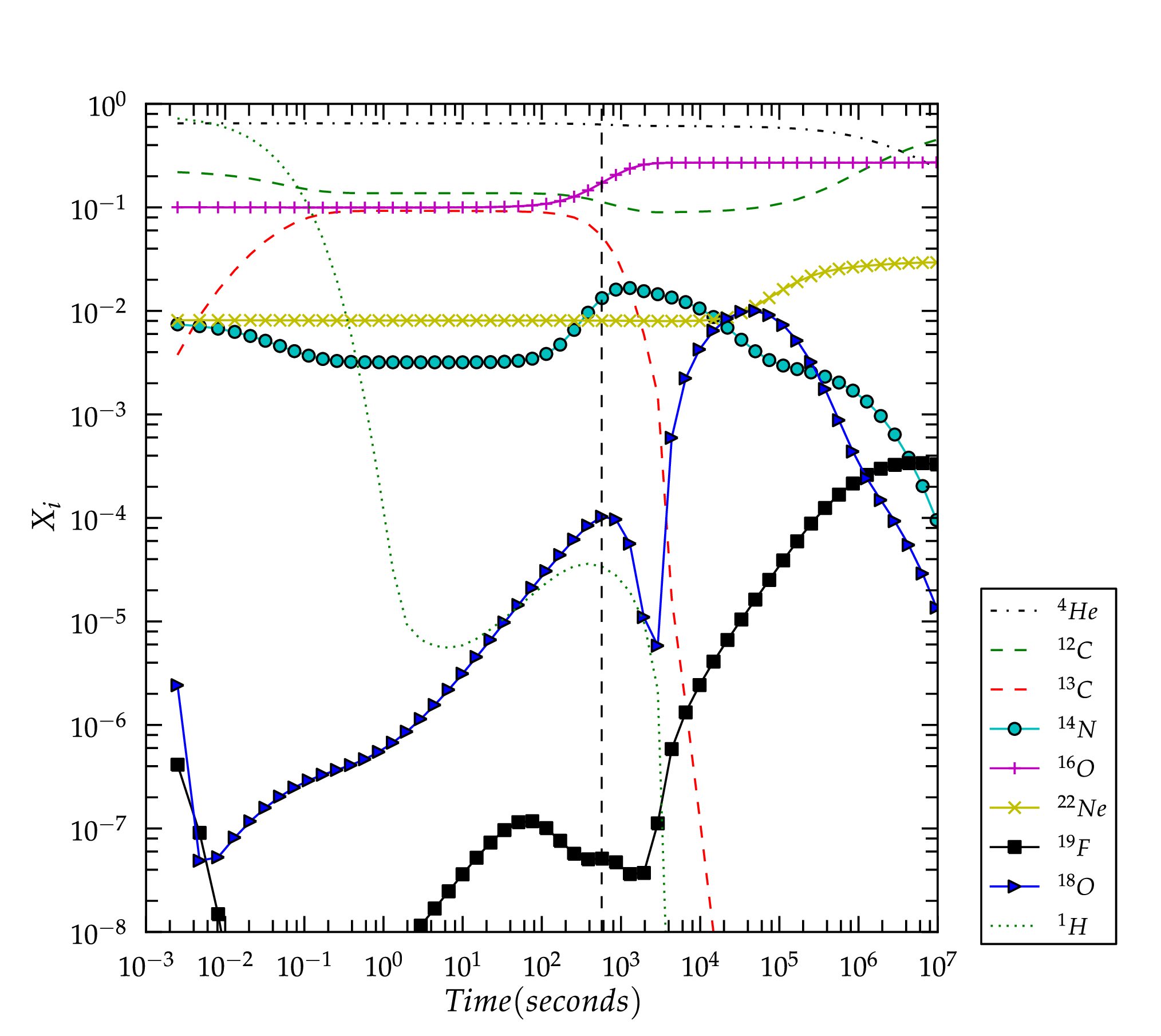}
\caption{Evolution of the nuclear species when starting with conditions
similar to those found in the SOF in the $q=0.5$ simulation with 
$T= 2.4 \times10^8K$ and $\rho=10^{4.5}$ over a period of $10^{6}$ seconds. The hydrogen abundance is multiplied by a factor of
$10^2$. The dashed vertical line corresponds to $\delta \rm t$
(Table~\ref{lowqdetailstable}). }
\label{img_0.5}
\end{figure}

\begin{figure}
\includegraphics[width=\textwidth]{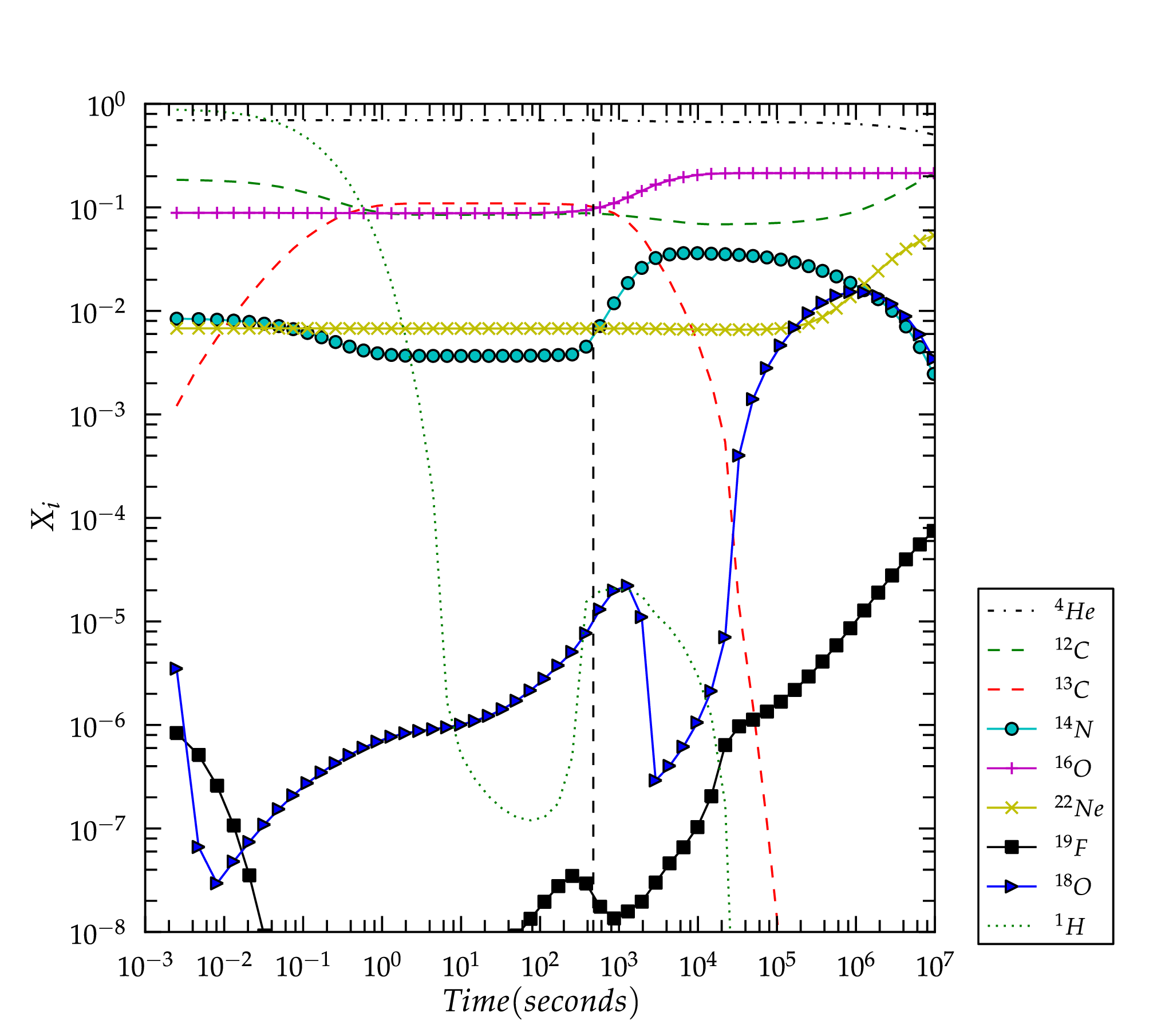}
\caption{Evolution of the nuclear species when starting with conditions
similar to those found in the SOF in the $q=0.6$ simulation at $T=
2.1\times10^8$ K and $\rho=10^{4.5}{\rm g/cm^3}$ over a period of $10^{6}$ seconds. The hydrogen abundance is multiplied by a factor of
$10^2$. The dashed vertical line corresponds to $\delta \rm t$
(Table~\ref{lowqdetailstable}).}
\label{img_0.6}
\end{figure}

\begin{figure}
\includegraphics[width=\textwidth]{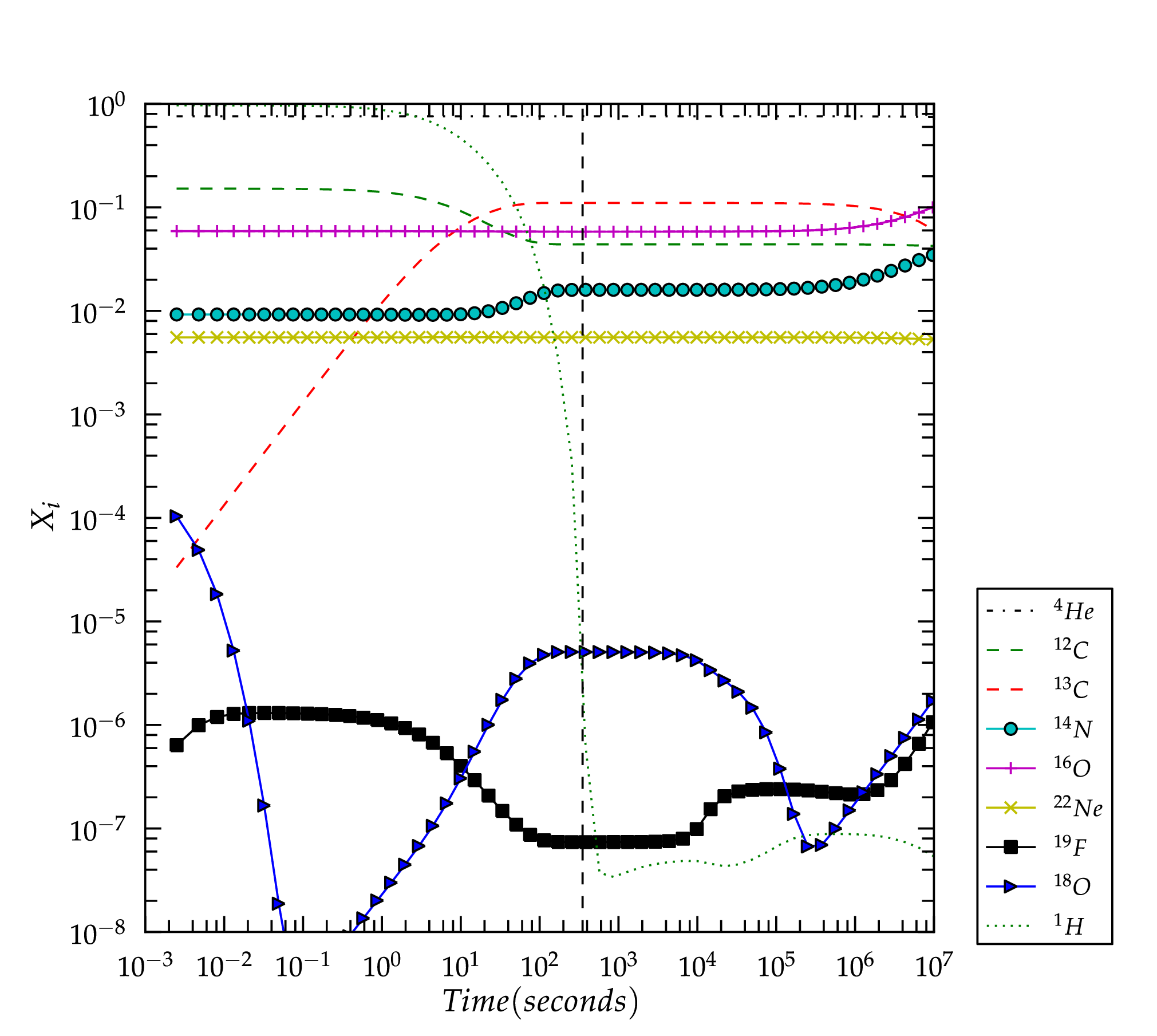}
\caption{Evolution of the nuclear species when starting with conditions
similar to those found in the SOF in the $q=0.7$ simulation at
$T=1.23\times10^8$ K and $\rho=10^{4.7} {\rm g/cm^3}$ over a period of $10^{6}$ seconds. The hydrogen abundance is multiplied by a factor of
$10^2$. The dashed vertical line corresponds to $\delta \rm t$ 
(Table~\ref{lowqdetailstable}).}
\label{img_0.7}
\end{figure}

The SOF is present until the end of the simulations for all three low-$q$
cases, hence it is not known for how long the conditions in the SOF would
last.  Therefore, we run the nucleosynthesis simulation until the abundance
of $^{18}\mathrm{O}$ begins to drop significantly.  The $^{18}\mathrm{O}$
abundance drops to $10^{-8}$ at $\sim 10^{7}$ seconds from the beginning of
the nucleosynthesis run.  We take this time period for the other two low-q
cases as well and run the nucleosynthesis network at the chosen constant
temperature and density (Table ~\ref{sofconditions}).  In order to compare
with observations, the abundances of all unstable elements are
instantaneously decayed.  These abundances are plotted in Figs. 
\ref{img_0.5}, \ref{img_0.6}, and \ref{img_0.7}, for $q=0.5$, $0.6$, and
$0.7$.  Since the densities in the SOF are similar amongst the low-$q$
cases, the differences in the final chemical abundances arise mainly due to
the different temperatures.  Hence in order to understand the role of
nuclear processes for different species, we take the case that showcases
them over the shortest amount of time, viz., the $q=0.5$ case.

In Fig.~\ref{img_0.5}, it is seen that between $10^{-3}$ and $10^{-1}$
seconds, proton capture reactions on $^{12}\mathrm{C}$, $^{14}\mathrm{N}$,
$^{18}\mathrm{O}$ and $^{19}\mathrm{F}$ bring down their initial abundances
by an order of 1.5 to 3.  Compared to the initial abundance, the
$^{18}\mathrm{O}$ abundance drops several orders of magnitude in $10^{-2}$
seconds.  This implies that the amount of initial $^{18}\mathrm{O}$
abundance does not help much in lowering the
$^{16}\mathrm{O}/^{18}\mathrm{O}$ ratio since most of the $^{18}\mathrm{O}$
present initially is destroyed.  The $^{13}\mathrm{N}$ abundance reaches a
quasi equilibrium value of 0.1 after 0.2 seconds via the destruction of
$^{12}\mathrm{C}$ by $^{12}\mathrm{C(p,\gamma)}^{13}\mathrm{N}$.  Since the
plotted abundances are only of stable nuclei, the rise in $^{13}\mathrm{N}$
abundance is reflected in the increase of $^{13}\mathrm{C}$.

From $10^{-1}$ to $10^{3}$ seconds, the above nuclei are regenerated by the
same proton capture reactions that caused their destruction earlier, viz.,
\begin{eqnarray}
^{17}\mathrm{O(p,\gamma)}^{18}\mathrm{F(^+\beta)}^{18}\mathrm{O} \nonumber\\
^{18}\mathrm{O(p,\gamma)}^{19}\mathrm{F}\\
^{19}\mathrm{F(p,\gamma)}^{20}\mathrm{Ne} \nonumber
\end{eqnarray}

During this time, the neutron abundance continually increases. The main
source of neutrons is the $^{13}\mathrm{C(\alpha,n)}^{16}\mathrm{O}$
reaction, along with $^{22}\mathrm{Ne(\alpha,n)}^{25}\mathrm{Mg}$ and 
other auxillary $(\alpha,n)$ reactions.  At $\sim$ 7 seconds, there is a rise 
in the proton abundance. The three main sources identified to cause an 
increase of 90$\%$ in the proton abundance are,
\begin{eqnarray}
^{15}\mathrm{O(n,p)}^{15}\mathrm{N} \nonumber\\
^{14}\mathrm{N(n,p)}^{14}\mathrm{C}\\
^{13}\mathrm{N(n,p)}^{13}\mathrm{C} \nonumber
\end{eqnarray}
along with smaller contributions from auxillary (n,p) reactions. 

The proton abundance begins to drop again at nearly 500 seconds as the
rapid consumption overwhelms the production.  At 1000 seconds, $\alpha$
capture on $^{13}\mathrm{C}$ becomes extremely efficient and its abundance
drops rapidly.


The neutron abundance reaches a peak at 2860 seconds and then drops quickly
due to
consumed by neutron capture reactions.  The protons are unable to increase
their abundance as the neutron and $^{14}\mathrm{O}$ abundance drop to
very low values.  At the same time that the proton abundance drops, the
$^{18}\mathrm{O}$ abundance drops suddenly due to
$^{18}\mathrm{O(p,\gamma)}^{19}\mathrm{F}$ and there is a simultaneous
increase in the $^{19}\mathrm{F}$ abundance.

Thereafter, it is the reign of partial helium burning reactions. Beginning from 3000 seconds, the abundance of $^{16}\mathrm{O}$,$^{18}\mathrm{O}$ and $^{19}\mathrm{F}$ increase via : 
\begin{eqnarray}
^{12}\mathrm{C(\alpha,\gamma)}^{16}\mathrm{O} \nonumber \\
^{14}\mathrm{N(\alpha,\gamma)}^{18}\mathrm{F(\beta^+)}^{18}\mathrm{O}\\
^{15}\mathrm{N(\alpha,\gamma)}^{19}\mathrm{F}\nonumber
\end{eqnarray}
At around 40,000 seconds $^{18}\mathrm{O}$ reaches its peak and begins to be 
converted to $^{22}\mathrm{Ne}$ by 
$^{18}\mathrm{O(\alpha,\gamma)}^{22}\mathrm{Ne}$. This destruction exceeds 
the production of $^{18}\mathrm{O}$ and its abundance drops to $10^{-8}$ at 
$10^{8}$ seconds and continues to drop as time goes on.

The $q=0.6$ and $0.7$ cases with a constant temperature of $T=1.23\times10^8K$ and
$T=2.1\times10^8 K$ (these temperatures are calculated assuming the abundances given in
Table~\ref{initial_abun}), respectively, show a much slower evolution of
nuclear abundances (Figs.~\ref{img_0.6}, and \ref{img_0.7}) compared to the
$q=0.5$ case.  Over a longer time period, these cases will also show the
same abundance trends as the $q=0.5$ case.

    
\begin{figure}
\includegraphics[width=\textwidth]{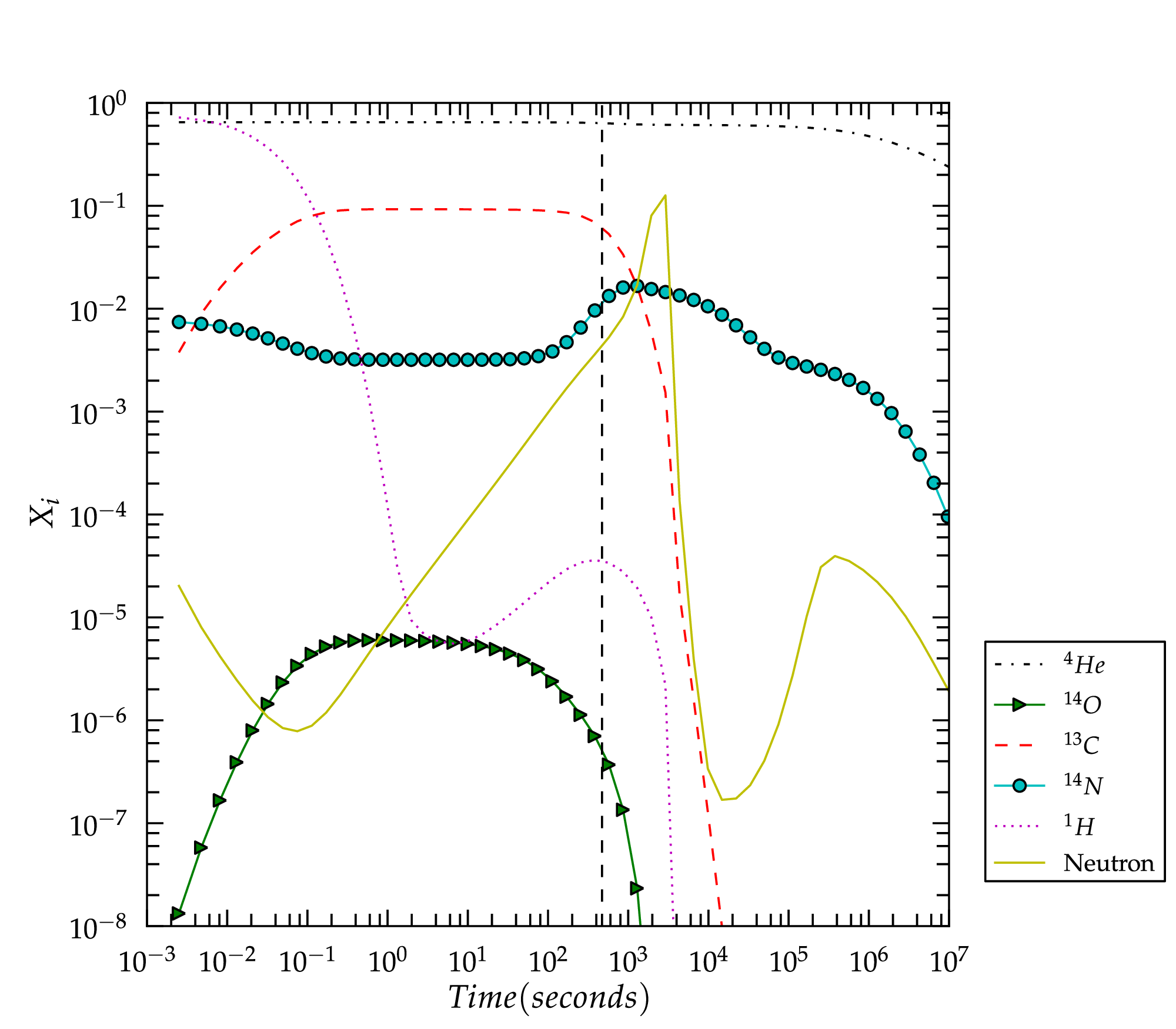}
\caption{Evolution of the neutron abundance in the SOF of the $q=0.5$ 
simulation, along with those species relevant to its evolution. The dashed 
vertical line corresponds to $\delta \rm t$ (Table~\ref{lowqdetailstable}). 
The neutron abundance is multiplied by a factor of $10^{13}$. }
\label{neutron_source}
\end{figure}

\begin{figure}
\includegraphics[width=\textwidth]{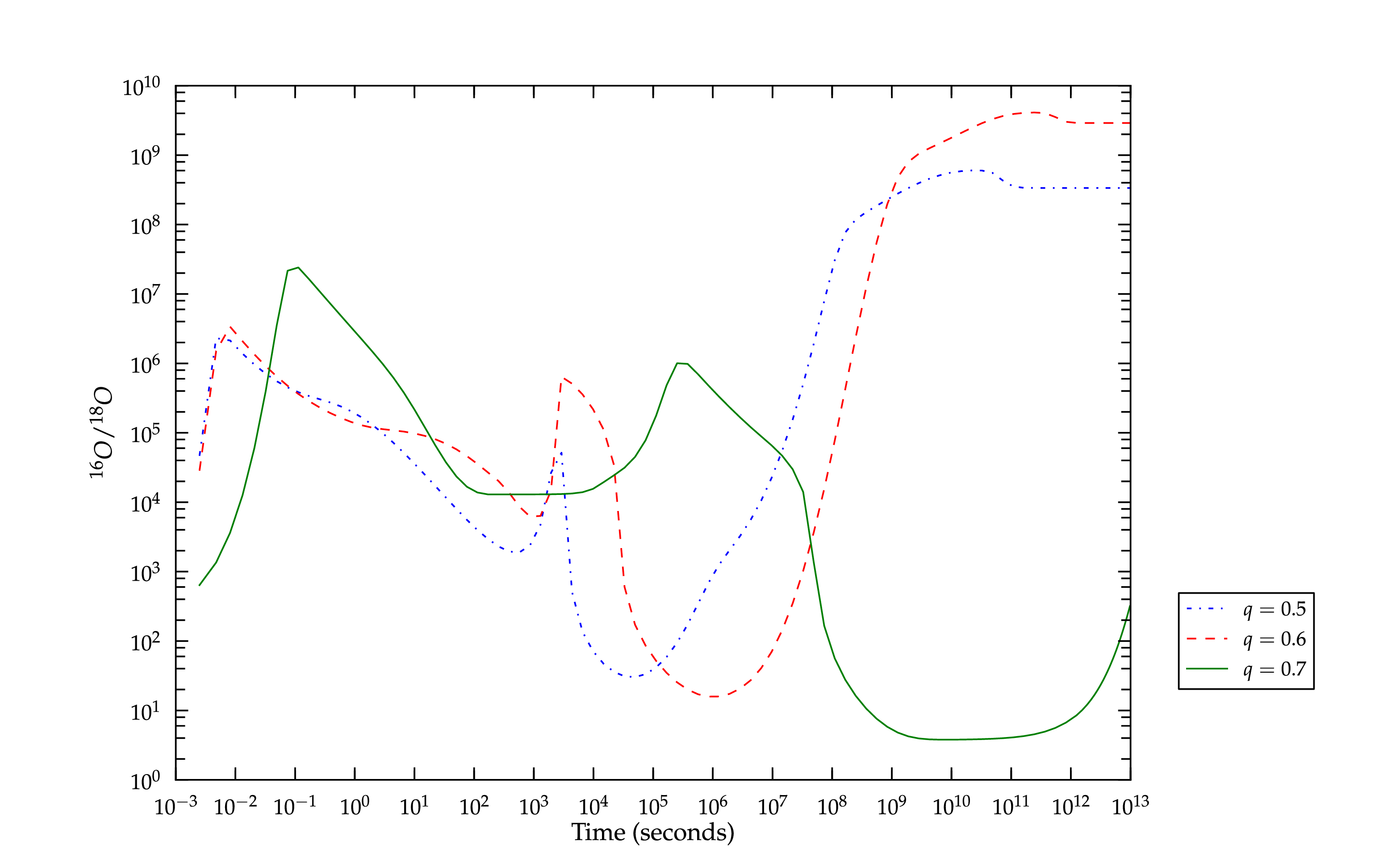}
\caption{Evolution of the $^{16}\mathrm{O}/^{18}\mathrm{O}$ ratio in the SOF of 
the $q=0.5$, $0.6$, and $0.7$ cases from $10^{-3}$ to $10^{13}$ seconds.}
\label{img_o16_o18}
\end{figure}

\begin{figure}
\includegraphics[width=\textwidth]{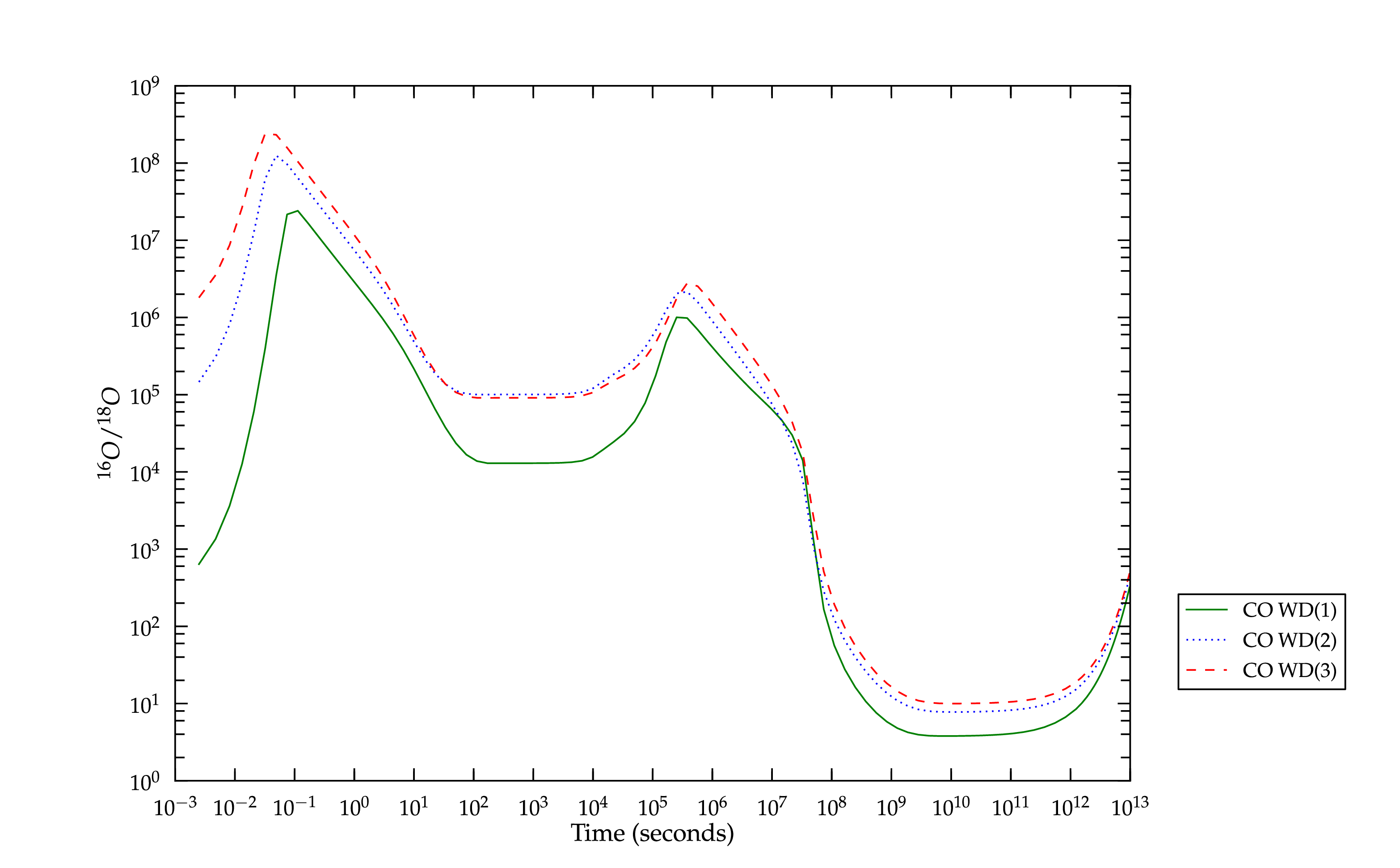}
\caption{Evolution of the $^{16}\mathrm{O}/^{18}\mathrm{O}$ ratio for 
initial abundances build from CO WD (1),(2) and (3) in Table~\ref{CO_WD_mod} for
the SOF of the $q=0.7$ case between $10^{-3}$ and $10^{13}$ seconds.}
\label{img_co_o16_o18}
\end{figure}

Fig.~\ref{img_o16_o18} plots the evolution of
$^{16}\mathrm{O}/^{18}\mathrm{O}$ in the SOF for each low-q case.  To
construct this plot the nucleosynthesis network was run for a longer time
period viz., $10^{13}$ seconds.  Within the time scale of the plots in
Figs.~\ref{img_0.5}, ~\ref{img_0.6}, and ~\ref{img_0.7} the lowest
$^{16}\mathrm{O}/^{18}\mathrm{O}$ value in the SOF is $\sim$ 16 and is
found in the $q=0.6$ case at about $10^{6}$ seconds after the SOF forms.  It
increases to 23 in $2.4{\times}10^{6}$ seconds and continues to increase
thereon.  In the $q=0.5$ case, the minimum value of this ratio is $\sim$
30, which is higher than the 0.6 case due to a faster destruction of
$^{18}\mathrm{O}$ to $^{22}\mathrm{Ne}$ while in the 0.7 case the minimum
value is ${\sim}$ 13,900 as there is hardly any $^{18}\mathrm{O}$ being
produced.

\begin{figure}
\includegraphics[width=\textwidth]{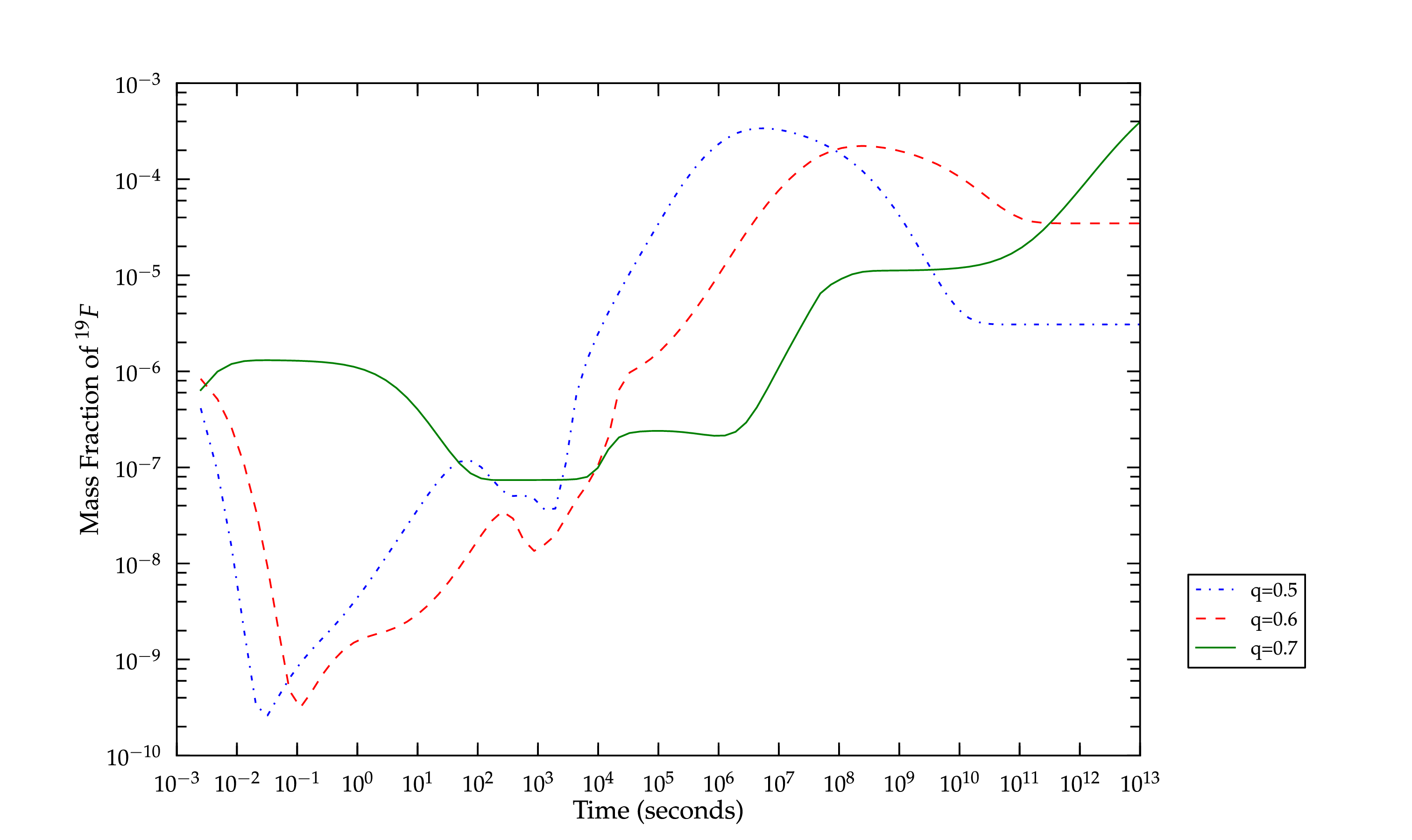}
\caption{Evolution of the $^{19}\mathrm{F}$ mass fraction in the SOF of 
the $q=0.5$, $0.6$, and $0.7$ cases from $10^{-3}$ to $10^{13}$ seconds.}
\label{f19_rcb}
\end{figure}

 Certain RCB stars are observed to be enriched in
$^{19}\mathrm{F}$ on their surface indicating an overabundance of
$^{18}\mathrm{O}$, however $^{18}\mathrm{O}$ and
$^{19}\mathrm{F}$ are not measured in the same stars. 
  The range of observed
$^{19}\mathrm{F}$ abundances in the RCB majority stars is between
$7.5\times10^{-6}$ to $3\times10^{-4}$ \citep{pandey08}. During the
course of the long term evolution of the $q=0.5$ case, the
$^{15}\mathrm{N(\alpha,\gamma)}^{19}\mathrm{F}$ reaction generates a high
amount of $^{19}\mathrm{F}$ and comes to an equilibrium value after $10^{10}$
seconds.  This amount falls within the range of the observed
$^{19}\mathrm{F}$ abundances of the majority of RCB stars for all three
low-q cases (Fig.~\ref{f19_rcb}). The $q=0.5$ and $0.6$ simulations show peaks
in the $^{19}\mathrm{F}$ abundance between $10^5$ and $10^9$ seconds.
Interestingly, the oxygen ratio is at a minimum of 16 at $10^6$ seconds for
$q=0.6$ (Fig.~\ref{img_o16_o18}, making it possible to have a low oxygen ratio at the same time as
$^{19}\mathrm{F}$ is enhanced.

However on maintaining the temperature, density conditions of the 0.7 case
for $10^{10}$ seconds, the lowest possible $^{16}\mathrm{O}/^{18}\mathrm{O}$
value in the SOF is found to be $\sim4$.  The oxygen isotopic ratio stays close
to 5 between $10^{9}$ and $10^{12}$ seconds.  Thus, it is possible to
reproduce in the SOF the $^{16}\mathrm{O}/^{18}\mathrm{O}$ ratio found in
RCB stars for a significantly large period of time.  However, it requires 
the initial abundances of the SOF of the 0.7 case and the sustenance
of constant temperature and density conditions of this case.

We recall that the maximum amount of $^{18}\mathrm{O}$
that can be produced is also limited by the amount of $^{14}\mathrm{N}$
present.  The metallicity of the He WD progenitor determines the initial
abundance of $^{14}\mathrm{N}$ but as can be seen in the nucleosynthesis
calculations, new $^{14}\mathrm{N}$ is formed by H-burning via the partial
CNO cycle.  The dredge up of accretor material and consequently the
$^{16}\mathrm{O}$ added to the SOF poses another constraint on this oxygen
isotopic ratio.

In order to confirm that this is the lowest
$^{16}\mathrm{O}/^{18}\mathrm{O}$ value one can get from our grid of CO WD
models (Sec.~\ref{progenitor_system}), the nucleosynthesis calculations are
also done by constructing initial abundances in the same manner as done
for the SOF of the $q=0.7$ case by using the same He WD model and the other
CO WD models, CO WD(2) and (3).  The evolution of the isotopic ratio for
these cases is shown in Fig.~\ref{img_co_o16_o18} and it can be seen that
indeed the lowest value is found for the progenitor system containing model
CO WD (1).

In order to estimate an approximate value of this ratio in the surface of
the merged object, we mix the post nucleosynthesis material of the SOF with
the layer above it.  The surface is defined as the SOF plus the layer on top
of it.  For the purpose of a rough estimate, we assume that the material
above the SOF has not undergone nucleosynthesis but just contains a
proportion of He and CO WD material according to Table~\ref{dredgeuptable}. 
For every timestep of the nucleosynthesis calculation of the SOFs of all
three low-q cases (as in Fig.~\ref{img_o16_o18}), we mix the abundances of
$^{16}\mathrm{O}$ and $^{18}\mathrm{O}$ with those in the unprocessed layer
above it.  The lowest value of $^{16}\mathrm{O}/^{18}\mathrm{O}$ amongst all
three cases thus obtained in the surface, is 4.6 (corresponding to the same
timestep at which 4 is obtained) belonging to the $q=0.7$ case.

The above results are from single zone nucleosynthesis calculations. It
would be important to perform multi-zone calculations as well in order to
consider the role of mixing between different layers of the star and its
consequences for the abundances.

\begin{figure}
\includegraphics[width=\textwidth]{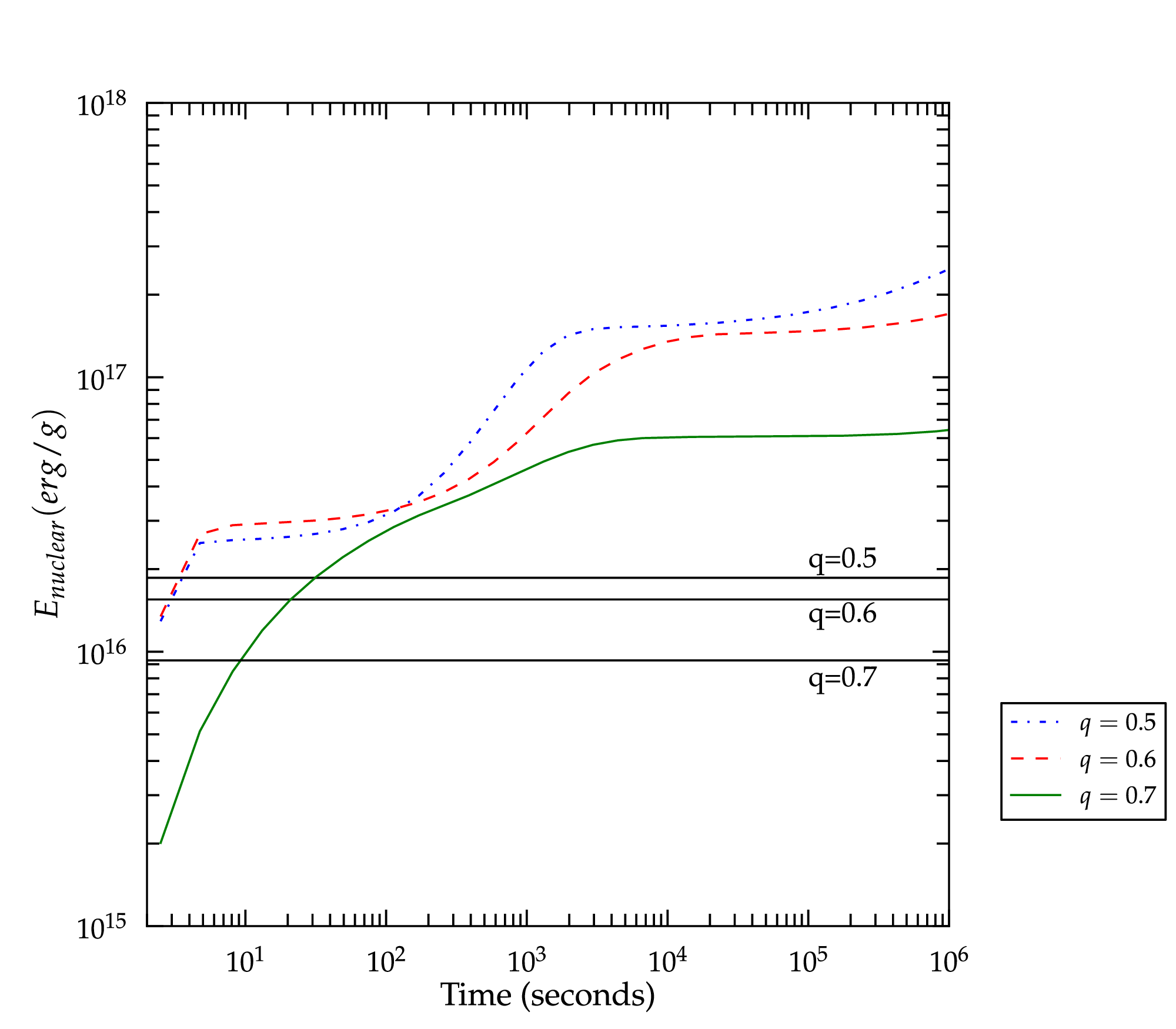}
\caption{The cumulative energy via nuclear processes released as a function of time
for the three cases. We note that these are upper limits on how much energy
would be added to the SOF, since 
energy lost by neutrinos are not taken into account, and hence may be higher 
by a factor
of 1.5-2. The solid horizontal black lines indicate the internal 
energy in the SOF at the end of the simulation.}
\label{Energy}
\end{figure}

From the nucleosynthesis calculations for each $q$, we also compute the
total energy released by nuclear processes per unit time in the SOF
(Fig.~\ref{Energy}).  Note that the energy calculated here does not account
for the loss of energy by neutrinos from weak interactions and therefore the
energy added to the SOF will be lower by a factor of 1.5-2.  Taking this
into account, we see from (Fig.~\ref{Energy}) that within the timescale of
the simulation ($\delta \rm t$ in Table~\ref{lowqdetailstable}) the nuclear
energy released is comparable to the internal energy of the SOF (depending
on the initial $q$).  The energy released in the first few hundred seconds
is mainly from proton captures.  Helium (for instance, triple $\alpha$ or
$^{14}\mathrm{N}(\alpha,\gamma)^{18}\mathrm{F}$) interacts on a much longer
timescale resulting in a plateau for $q=0.5$ and $0.6$ in Fig.~\ref{Energy}
between 10 and 100 seconds.  The extra energy released from nuclear
processes may lead to higher temperatures, and these processes could play
an important role in determining the temperature of the SOF.
Cooling processes and dynamical effects may also be
important but these cannot be estimated with our current tools.

\section{Comparison to recent WD merger simulations}

Our nucleosynthesis calculations were made assuming constant temperature and
density.  In the SOF, the temperature in the hydrodynamics simulations are
found to fluctuate significantly, especially during the actual merger where
the temperature may briefly reach up towards $3.5\times10^8$ K (assuming C and O
only).  Also, the SOF spans a range of densities, something that our
nucleosynthesis calculations cannot capture.  More elaborate nucleosynthesis
calculations (preferably included in the hydrodynamics simulations) should
be performed in order to find accurate abundance ratios.  
The simulation discussed in \citet{longland11}
does include a (simple) nucleosynthesis network in the dynamics
calculations followed by a more elaborate nucleosynthesis calculation in the
post processing, and they find  $^{16}\mathrm{O}$ to
$^{18}\mathrm{O}$ ratios similar to this study.

The oxygen isotopic ratio is $\sim1200$ in the $q=0.5$ simulation reported
here after 1000
seconds (the dynamical time scale), and is much higher than the lowest value
of 19 reported for $q=0.5$ in \citet{longland11}, and about the same as their ratio of
370 for a fully convective envelope.  
The total mass in that
simulation is higher than the total mass used in our simulations which will
affect the temperatures and maybe also the dredge-up, possibly explaining
the differences between their results and ours.

\citet{raskin11} also presented simulations of WD mergers
with higher total masses (their focus was on type Ia supernovae).  A
noticeable difference between their findings and ours is that they have an
``SOF'' (or at least a hot ring surrounding their merged core; the 3D
temperature structure is not shown in the paper) even in the equal mass
simulations.  We only find a post merger SOF for $q\lesssim0.7$.  The merged
core in the equal mass ($q=1$) simulations in \citet{raskin11} appears to be cold,
indicating that there has not been much mixing occurring in the core (however
in their $0.64 M_\odot \times 2$ simulation the cores appear to mix
similarly to our findings; this is also their simulation that is closest to
ours in mass). 
The SPH simulations of \citet{longland11} and \citet{raskin11} are not only
for a different total mass, but also include
nucleosynthesis, and it is plausible that it can cause the differences between
their results and ours,
especially the fairly rapid burning of the helium atmosphere of their WDs. 
Our hydrodynamics simulations did not include the energy released from nuclear
processes. We showed that this energy is comparable to the thermal energy in
the SOF. In the short period of time that the hydrodynamics simulations ran,
this extra energy input will probably not make much of a difference.
However, on a longer time scale, with even more energy input from nuclear
reactions, it is possible that the conditions in the SOF will not remain
constant as we assumed in Fig.~\ref{img_o16_o18}. However, a higher
thermal energy can also make the SOF expand.
Following the dynamics for a much longer time, including the energy produced
in nuclear processes, is needed in order to fully understand the evolution
of the SOF.

\section{Discussion}

The ratio of $^{16}{\mathrm O}$ to $^{18}{\mathrm O}$ is observed to be very
low (of the order unity) in RCB stars.  From hydrodynamic simulations of
double degenerate
mergers for various mass ratios, we have therefore looked for conditions
that would allow for production of $^{18}{\mathrm O}$ in order to explain
this ratio.  This requires temperatures of the order $\sim1.5$ -
$3\times10^8$ K.  At
lower temperatures, $^{18}{\mathrm O}$ will not form on the available
dynamic time scale, and at higher temperatures it will be destroyed
(converted to $^{22}\mathrm{Ne}$) as soon as it forms.  Our results show
that the maximum temperature that can be found in the SOF depends strongly
on the mass ratio, $q$.  Low values of $q$ give temperatures in the SOF of
$1.2-2.5 \times10^8$ K (assuming realistic abundances), while mass ratios above
$q=0.7$ gives temperatures much lower than that.  Hence the lower $q$
values that we have investigated will give temperatures in the SOF suitable
for $^{18}{\mathrm O}$ production on a dynamical timescale.

On a dynamical time scale, we do not find very low oxygen ratios in the SOF
in any of our simulations.  After a thousand seconds, the $q=0.5$ simulation
reaches a $^{16}\mathrm{O}$ to $^{18}\mathrm{O}$ ratio of about 2000, since not
much $^{18}\mathrm{O}$ is produced on such a short time scale and because of
the large amount of $^{16}\mathrm{O}$ present from the dredge up.  Within a
day, the oxygen ratio in the $q=0.5$ simulation reaches its lowest value of
$\sim30$
(Fig.~\ref{img_o16_o18}). After that, $^{18}\mathrm{O}$ is being destroyed
and the ratio increases.  The $q=0.6$ simulation reaches its lowest value of
16 after $10^6-10^7$ seconds, after which the ratio increases again as
$^{18}\mathrm{O}$ is being destroyed.  The best case for obtaining low
oxygen ratios is the $q=0.7$ simulation, which reaches the lowest
$^{16}\mathrm{O}$ to $^{18}\mathrm{O}$ of $4$ after $\sim10^{2}$ years, assuming
the conditions remain constant for that long.  This is comparable to the
observed oxygen ratios in RCB stars. 
 As we have shown, the nuclear reactions begin very fast, and the extra 
energy released is likely to affect the
conditions in the SOF. Unfortunately we cannot model this at present, and
here we simply state that if the very low oxygen ratios of order unity shall be
achieved in the $q=0.7$ simulations, the conditions in the SOF must remain relatively unchanged for a
period of about a hundred years.
 
In the high-$q$ simulations the temperature was not
sufficiently high to produce  $^{18}\mathrm{O}$. However, 
as we have shown the protons react very quickly,
releasing an amount of energy comparable to the thermal energy in the SOF. 
Hence, it is possible that even in the high $q$ simulations, the SOF prior
to the merger can become sufficiently hot for the nucleosynthesis involving
helium to start.  This extra energy from nuclear processes leads to an extra
pressure term, that could potentially help preserve the SOF also in the
high-$q$ simulations so that the $^{18}\mathrm{O}$ production can continue. 
Furthermore, even though the amount of $^{18}\mathrm{O}$ in RCBs is strongly
enhanced compared to all other known objects, it is important to keep in
mind that among RCB stars the observed oxygen ratio varies by 2 orders of
magnitude from star to star.

The $^{16}{\mathrm O}$ to $^{18}{\mathrm O}$ ratio depends both on the
formation of $^{18}{\mathrm O}$, and also on the amount of $^{16}{\mathrm
O}$ present.  In all of our simulations, we have found
significant dredge up of accretor material which consists primarily of
$^{16}{\mathrm O}$ and $^{12}{\mathrm C}$. $^{18}{\mathrm O}$ is formed from  $^{14}{\mathrm N}$, which in
part depends on the initial metallicity of the progenitor stars,
assumed to be solar giving about $1\%$ of $^{14}{\mathrm N}$. But
$^{14}\mathrm{N}$ is also being produced
(Figs.~\ref{img_0.5}, \ref{img_0.6}, and \ref{img_0.7}) in the nuclear
processes occurring in the SOF. The $^{16}\mathrm{O}$ to
$^{18}\mathrm{O}$ ratios, that we find in the SOF (shown in
Figs.~\ref{img_o16_o18}),  are the lowest
values possible from our simulations.

If less $^{16}\mathrm{O}$ were dredged up from the accretor, then less
$^{18}\mathrm{O}$ would need to be produced in order to obtain low oxygen
ratios.  One way to avoid contaminating the SOF with $^{16}\mathrm{O}$ from
dredge-up, would be if the accretor is a hybrid He/CO WD.  \citet{rappaport09}
modeled a $0.475 M_\odot$ hybrid He/CO WD with a He envelope of more than $0.1
M_\odot$.  In our simulations, we found that about $0.1 M_\odot$ of accretor
material was dredged-up, most of it ending up in the SOF.  If the accretor
is a hybrid He/CO WD, most of this dredged-up material might turn out to be
$^4\mathrm{He}$, and the $^{16}\mathrm{O}$ contamination of the SOF would be
much less.  In this case, the lower mass of the accretor means the donor must also have a
lower mass in order to get a tidal disruption of the donor rather than a
core merging.  This lower total mass could result in lower
temperatures.  Furthermore, a mixture with mostly He could also
lead to lower temperatures (Eqs.~\ref{eq:T} and \ref{eq:cv}).  However, as
we discussed above, the reacting protons might heat the gas to a sufficiently
high temperature to start the helium burning no matter what the total mass. 
Hence, it remains to be seen if the oxygen isotope ratio will be of the
correct order if this is the situation.  
\citet{han02, han03} showed that a significant fraction of sdB stars are
expected to be in close-period binaries with He WDs.  After completion of
helium burning, sdB stars may evolve into hybrid CO/He WDs \citep{justham11}. 
In \citet{clausen12}, a population synthesis study found that there should be
almost 200,000 sdB stars in binaries in the galaxy.  Hence it may not be
unreasonable that a small fraction of these are in a close binary with a He
WD, and that after the sdB star has evolved into a hybrid WD they can merge.


When the merged object begins to expand, the temperature and density of the
SOF are likely to drop which may bring the nuclear processes there to a halt. 
Hence the oxygen ratio in the SOF at that time may be
representative of what will be observed at the surface of this object. 
With our numerical approach, we cannot determine when the merged object will
begin to expand.

Convection may be triggered quickly in the SOF, as the thermal diffusion time
\citep[$\tau_{\rm th}\sim10^9$ seconds using $\varepsilon_{\rm th}\sim 10^7
{\rm ~erg~g^{-1}~s^{-1}}$ from][]{yoon07} is much larger than the 100 seconds
thermonuclear time scale ($\tau_{\rm nuc}=c_p T/\epsilon_{\rm nuc}$, where
$c_p$ is the specific heat at constant pressure that we estimate to be of
order $10^8$ erg/g K, $T$ is the temperature of order $10^8$ K, and
$\epsilon_{\rm nuc}$ is the nuclear energy production rate of order $10^{14}
{\rm ~erg~g^{-1}~s^{-1}}$ after 1000 seconds; Fig.\ref{Energy}).

If the material outside the merged core and the SOF accretes quickly, it
could affect the SOF.  \citet{ken12} argued that
the viscous time scale for material in a disk surrounding such a core is an hour to a year, indicating
that it could accrete relatively fast.
Alternatively, the merged object can expand before this material has
accreted.  In that case, the SOF is unlikely to be affected much by this
material.  However, the \citet{ken12} result would indicate that the
expansion would have to begin quickly (less than a year) since the accretion may occur
on this time scale. We found
in the $q=0.6$ simulation that the oxygen ratio is 16 after only $10^6$ seconds,
a significant enhancement from the solar value and comparable to what is
seen in some RCBs.

We note that most of the accretor material, that is being
dredged up, ends up in the SOF.  The material outside the SOF is mostly from
the He WD (Table~\ref{dredgeuptable}).  The conditions outside the SOF are
not favorable for nuclear processes to occur, and so if only
the matter outside the SOF
swells up to supergiant size (without mixing in newly produced elements in
the SOF), the resulting object could look like a helium star with some
carbon, ie. an RCB star.

We have found that the best conditions for
reaching the low oxygen ratios occur if the temperature remains 
$\sim10^8$ K.  Then $^{18}\mathrm{O}$ forms slowly,
and is destroyed even more slowly.  Hence it is plausible that
the merger of a CO WD and a He WD can lead to the formation of an RCB star,
although a more elaborate investigation must be performed in order to
conclusively answer this question.

\acknowledgements
We thank J. Frank for helpful discussions and comments.
This work has been supported, in part, by grants AST-0708551, OIA-0963375,
and PHY0922648 (JINA) from the U.S. National Science Foundation and, in
part, by grant NNX10AC72G from NASA's ATP program.
The NuGrid collaboration acknowledges support from the Joint Institute for
Nuclear Astrophysics (JINA, supported by NSF grant PHY0922648). FH
acknowledges funding through an NSERC Discovery Grant.
Work at LANL was done under the auspices of the National Nuclear Security
Administration of the
U.S. Department of Energy at Los Alamos National Laboratory under Contract
No. DE-AC52-06NA25396.
MP also thanks support from the Ambizione grant of the SNSF (Switzerland),
and from the European research program Eurogenesis.
The computations have been performed in part on the LONI machine Queenbee
through grant loni lsuastro09, the Teragrid machines Abe and Kraken through
grant TG-AST090104, and the LANL Institutional Computing machine Lobo. 

%
%

\end{document}